%% file: starck.tex
\documentclass[twocolumn,traditabstract]{aa}
\usepackage{natbib}
\usepackage[english]{babel}
\usepackage{algorithmic}
\usepackage{algorithm}
\usepackage{theorem}
\usepackage{amssymb}
\usepackage{amsmath}
\usepackage{mathtools}
\usepackage{graphics}
\usepackage{subfigure}
\usepackage{fullpage}
\usepackage{url}
\usepackage{hyperref,enumerate}
\usepackage{epsfig}
\newcommand{\bibfont}{\tiny}

\newcommand{\cmb}{{\boldsymbol{x}}}
\newcommand{\cmba}{{\boldsymbol{a}}}
\newcommand{\obs}{{\boldsymbol{z}}}
\newcommand{\mask}{{\mathbf{M}}}
\newcommand{\sht}{{\mathbf{S}}}
\newcommand{\fsky}{{\mathrm{Fsky}}}

\newcounter{mycomment}

\graphicspath{./figures/} 

\usepackage{amstext,amssymb,amsbsy,amsmath,amscd,mathrsfs,dsfont}
\usepackage{graphics,graphicx,epsfig,subfigure,enumerate,rotating,theorem,algorithm,algorithmic,array,longtable}

\include{prelim}

\begin{document}
\title{Low-$\ell$ CMB Analysis and Inpainting}

\author{J.-L. Starck \inst{1} and M.J. Fadili \inst{2} and A. Rassat \inst{3,1}}

\institute{
\inst{1} Laboratoire AIM, UMR CEA-CNRS-Paris 7, Irfu, SAp/SEDI, Service d'Astrophysique, CEA Saclay, F-91191 GIF-SUR-YVETTE Cedex, France  \\
\inst{2} GREYC CNRS-ENSICAEN-Universit\'e de Caen, 6, Bd du Mar\'echal Juin, 14050 Caen Cedex, France\\
\inst{3} Laboratoire d'Astrophysique, Ecole Polytechnique F\'ed\'erale de Lausanne (EPFL), Observatoire de Sauverny, CH-1290, Versoix, Switzerland \\
}
 
\date\today

\abstract{Reconstruction of the CMB in the Galactic plane is extremely difficult due to the dominant foreground emissions such as Dust, Free-Free or Synchrotron. 
For cosmological studies, the standard approach consists in masking this area where the reconstruction is not good enough. This leads to difficulties for  
the statistical analysis of the CMB map, especially at very large scales (to study for e.g., the low quadrupole, ISW, axis of evil, etc). 
We investigate in this paper how well some inpainting techniques can recover the low-$\ell$ spherical harmonic coefficients. 
We introduce three new inpainting techniques based on three different kinds of priors: sparsity, energy and isotropy, and we compare them.
We show that two of them, sparsity and energy priors, can lead to extremely high quality reconstruction, within 1\% of the cosmic variance for
a mask with Fsky larger than 80\%.
}
 

\maketitle 
 
 
\section{Introduction}
\subsection{CMB and the masked data problem}
As component separation methods do not provide good estimates of the Cosmic Microwave Background (CMB)
in the Galactic plane and at locations of point sources, the standard approach for CMB map analysis is to consider  that the data are not reliable in these areas, and to mask them. This leads to an incomplete coverage of the sky that has to be handled properly for further analysis. This is especially true for analysis methods which operate in the spherical harmonic domain where localization is lost and full-sky coverage is assumed. 
For power spectrum estimation, methods like MASTER \citep{master:hivon02} solve a linear ill-posed inverse problem allowing to deconvolve the observed power spectrum of the masked map from the mask effect. 

For non-Gaussianity analysis, many approaches have been proposed to deal with this missing data problem.
Methods to solve this problems are called gap filling or inpainting methods.

{ The simplest approach is to set pixel values in the masked area to zero. This is sometimes claimed to be a good approach by assuming it does not add any information. However, this is not correct, as setting the masked area to zero actually adds (or removes) information, which are also not the true CMB values.} 
A consequence of this gap filling technique is the creation of a lot of artifactual coefficients at the border of the mask in a wavelet analysis or a leakage between different multipoles in a spherical harmonic analysis. This effect can be reduced using an apodized mask. A slightly more sophisticated method consists in replacing each missing pixel by the average of its neighbors and iterating until the gaps are filled. This technique is called diffuse inpainting and has been used in Planck-LFI data pre-processing \cite{lfi2011}. It is acceptable for treatment of point sources, but is far from being a reasonable solution for the Galactic plane inpainting in CMB non-Gaussianity analysis. 

In \citet{inpainting:abrial06,starck:abrial08}, the problem was considered as an ill-posed inverse problem, $\obs = \mask \cmb$, where $\cmb$ is the unknown CMB, $\mask$ is the masking operator, and $\obs$ the masked CMB map. A sparsity prior
in the spherical harmonic domain was used to regularize the problem. This sparsity-based inpainting approach has been successfully used for two different CMB studies, the CMB weak lensing on 
Planck simulated data \citep{perotto10,stephane2012},
and the analysis of integrated Sachs-Wolfe effect (ISW) on WMAP data  \citep{starck:dupe2011}. In both cases, it was shown using Monte-Carlo simulations that the statistics derived from the inpainted maps can be trusted with high confidence level, and that sparsity-based inpainting can indeed provide an easy and effective solution to the large Galactic mask problem. 

It was also shown that sparse inpainting is useful for weak lensing data \citep{starck:pires08}, Fermi data \citep{starck:schmitt2010}, and asteroseismic data \citep{sato2010}.
The sparse inpainting success has motivated the community to investigate more inpainting techniques, and
other approaches have been recently proposed.  

\citet{bucher2012,kim2012} seek a solution which complies with the CMB properties, i.e. to be a homogeneous Gaussian random field with a specific power spectrum. Nice results were derived, but the approach presents however the drawback
that we need to assume a given cosmology, which may not be wise for non-Gaussianity studies. 
For large scale CMB anomalies studies, \citet{peiris2011,david2012}, building upon the work of \citet{OliveiraTegmark06} proposed to use generalized least-squares (GLS), which coincide with the maximum likelihood estimator (MLE) under appropriate assumptions such as Gaussianity. This method also requires to have the input power spectrum. 
In addition, when the covariance in the GLS is singular, the authors propose to regularize it by adding a small perturbation term that must also to ensure positive-definiteness. Using this regularization, the estimator is no longer a GLS (nor a MLE). 

\subsection*{Which model for the CMB ?}

As the missing data problem is ill-posed, prior knowledge is needed to reduce the space of candidate solutions.  
Methods attacking this problem in the literature assume priors, either explicitly or implicitly. 
If we put aside the zero-inpainting and the diffuse inpainting methods which are of little interest for the Galactic plane inpainting, the two other priors were:
\begin{enumerate}[$\bullet$]
\item{\bf Gaussianity:} the CMB is assumed to be a homogeneous and isotropic Gaussian random field, so it makes sense to use such a prior. In practice, methods require the theoretical power spectrum, and it has either to be estimated using a method like TOUSI \citep{paniez:tousi}, or a cosmology has to be assumed, which is even a stronger assumption than the Gaussianity prior. We should also keep in mind that the goal of non-Gaussianity studies is to check that the observed CMB does not deviate from Gaussianity. Therefore, we should be careful with this prior. 
\item{\bf Sparsity:} the sparsity prior on a signal consists in assuming that its coefficients in a given representation domain, when sorted in decreasing order of magnitude, exhibit a fast decay rate, typically a polynomial decay with an exponent that depends on the regularity of the signal. For the CMB, its spherical harmonic coefficients show a decay in $O(\ell^{-2})$ up to $\ell$ around $900$ and then $O(\ell^{-3})$ for larger multipoles. Thus, the sparsity prior is advocated, and this explains its success for CMB-related inverse problems such as inpainting or component separation \citep{bobin2012}.
\end{enumerate}

\subsection{Contributions}
In this paper, we revisit the Gaussianity and sparsity priors, and introduce an additional one, namely the CMB isotropy, for the the recovery of the spherical harmonic coefficients at low $\ell$ ($<10$) from masked data. We describe novel and fast algorithms to solve the optimization problems corresponding to each prior. These algorithms originate from the field of non-smooth optimization theory and they apply efficiently to large-scale data. We then show that some of these inpainting algorithms are very efficient to recover the spherical harmonic coefficients for $\ell < 10$ when using the sparsity or energy priors. We also study the quality the reconstruction as a function of the sky coverage, and we show that a very good reconstruction quality, within 1\% of the cosmic variance, can be reached for a mask with sky coverage larger than 80\%.



\section{CMB Inpainting}

\subsection{Problem statement}
Suppose that we observe $\obs = \mask \cmb$, where $\cmb$ is a real-valued centered and square-integrable field on the unit sphere $\mathbb{S}^2$, and $\mask$ is a linear bounded masking operator. The goal is to recover $\cmb$ from $\obs$.

The field $\cmb$ can be expanded as
\begin{align*}
\cmb(p)=\sum_{\ell=0}^{+\infty}\sum_{m=-\ell}^{\ell} \cmba_{\ell,m}Y_{\ell m}(p)~, \\
\text{ where } \cmba_{\ell,m} = \int_{\mathbb{S}^2} \cmb(p) Y^*_{\ell m}(p) dp~,
\end{align*}
where the complex-valued functions $Y_{\ell m}$ are the co-called spherical harmonics, $\ell$ is the multipole moment and $m$ is the phase ranging from $-\ell$ to $\ell$. The $\cmba_{\ell,m}$ are the spherical harmonic coefficients of $\cmb$. In the following we will denote $\sht$ the spherical harmonic transform operator and $\sht^*$ its adjoint (hence its inverse since spherical harmonics form an orthobasis of $\mathbb{L}_2(\mathbb{S}^2)$).

If $\cmb$ is a wide-sense stationary (i.e. homogeneous) random field, the spherical harmonic coefficients are uncorrelated, 
\begin{equation*}
\E{\cmba_{\ell,m}^*\cmba_{\ell^\prime,m^\prime}} = \delta_{\ell\ell^{\prime}}\delta_{mm^{\prime}}C_{\ell,m} ~.
\end{equation*}
If moreover the process is isotropic, then
\begin{equation*}
\label{eq:almvar}
\E{|\cmba_{\ell,m}|^2} = C_{\ell} ~, \quad -\ell \leq m \leq \ell ~,
\end{equation*}
where $C_{\ell}$ is the angular power spectrum, which depends solely on $\ell$. 

In the rest of the paper, we consider a finite-dimensional setting, where the sphere is discretized. $\cmb$ can be rearranged in a column vector in $\RR^n$, and similarly for $\obs \in \RR^m$, with $m < n$, and $\cmba \in \CC^p$. Therefore, $\mask$ can be seen as a matrix taking values in $\{0,1\}$, i.e. $\mask \in \mathscr{M}_{m \times n}(\{0,1\})$. The goal of inpainting is to recover $\cmb$ from $\obs$.
  
\subsection{A General Inpainting Framework}
The recovery of $\cmb$ from $\obs$ when $m < n$ amounts to solving an underdetermined system of linear equations. Traditional mathematical reasoning, in fact the fundamental theorem of linear algebra, tells us not to attempt this: there are more unknowns than equations. However, if we have prior information about the underlying field, there are some rigorous results showing that such inversion might be possible \citep{starck:book10}.

In general, we can write the inpainting problem as the following optimization program
\begin{equation}
\label{eq:mingen}
\text{find } \widehat{\cmb} \in \Argmin{\cmb \in \RR^n} R(\cmb) \st \obs-\mask \cmb \in \cC ~,
\end{equation}
where $R$ is a proper lower-bounded penalty function reflecting some prior on $\cmb$, and $\cC$ is closed constraint set expressing the fidelity term. Typically, in the noiseless case, $\cC=\{\cmb: \obs = \mask \cmb\}$. This is the situation we are going to focus on in the rest of the paper. As far as $R$ is concerned, we will consider three types of priors, each corresponding to a specific choice of $R$.

\section{Sparsity Prior}
\label{sec:sparsity}

Sparsity-based inpainting has been proposed for the CMB in \cite{inpainting:abrial06,starck:abrial08} and for weak-lensing mass map reconstruction in \cite{starck:pires08,pires10}. In \citet{perotto10,starck:dupe2011,starck:rassat2012}, it was shown that sparsity-based inpainting does not destroy CMB  weak-lensing, ISW signals {nor some large scale anomalies in the CMB}, and is therefore an elegant way to handle the masking problem. Note that the masking effect can be thought of as a loss of sparsity in the spherical harmonic domain because the information required to define the map has been spread across the spherical harmonic basis (leakage effect). 


\subsection*{Optimization problem}
If the spherical harmonic coefficients $\cmba$ of $\cmb$ (i.e. $\cmba =\sht \cmb$) are assumed to be sparse, then, a well-known penalty to promote sparsity is the $l_q$ (pseudo- or quasi-)norm, with $q \in [0,1]$. Therefore, \eqref{eq:mingen} becomes
\begin{equation}
\label{eq:minsparse}
\text{find } \widehat{\cmba} \in \Argmin{\cmba \in \CC^p} \norm{\cmba}^q_q \st \obs = \mask {\sht}^*  \cmba ~,
\end{equation}
where $\norm{\cmba}_q^q=\sum_{i} \abs{\cmba_i}^q$, and where $\abs{z}=\sqrt{\Re(z)^2+\Im(z)^2}$ for $z\in\CC$. For $q=0$, the $l_0$ pseudo-norm counts the number of non-zero entries of its argument. The inpainted map is finally reconstructed as $\widehat{\cmb}= {\sht}^* \widehat{\cmba}$.



Solving \eqref{eq:minsparse} when $q=0$ is known to be NP-hard. This is further complicated by the presence of the non-smooth constraint term.
 Iterative Hard thresholding (IHT), described in Appendix A, attempts to solve this problem. 
 It is also well-known that $l_1$ norm is the tightest convex relaxation (in the $\ell_2$ ball) of the $l_0$ penalty \citep{starck:book10}. 
 This suggests solving \eqref{eq:minsparse} with $q=1$. In this case, the problem is well-posed: it has at least a minimizer and all minimizers are global. 
 Furthermore, although it is completely non-smooth, it can be solved efficiently with a provably convergent algorithm belonging to the family of 
 proximal splitting schemes \citep{CombettesPesquet09,starck:book10}. This can be done using the Douglas-Rachford (DR) algorithm described in Appendix B.


\subsection*{Comparison between IHT and DR}

\begin{figure*}[htb]
\centering
\includegraphics [trim= 3cm 13cm 2cm  2.6cm , clip,  width=\textwidth]{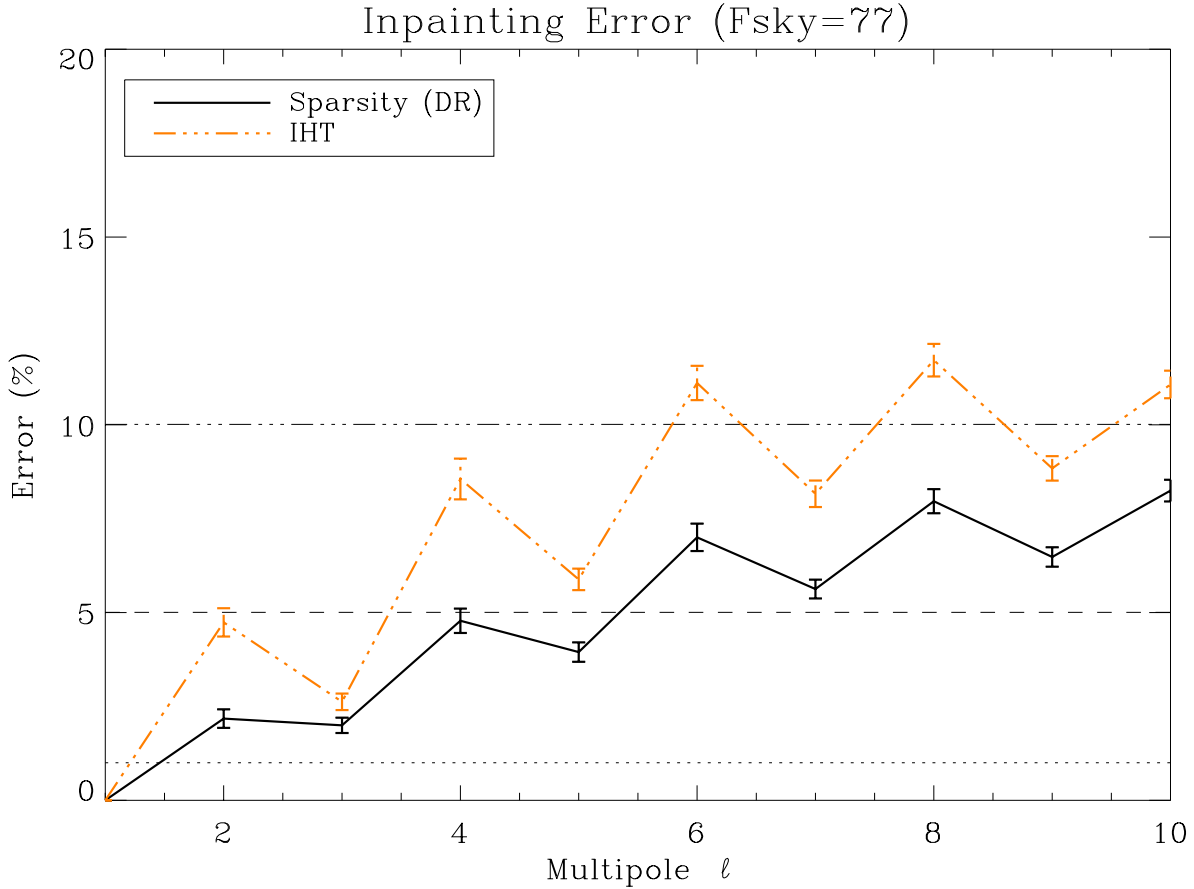}
\caption{Relative MSE per $ \ell $ in percent for the two sparsity-based inpainting algorithms, IHT (dashed blue line) 
and DR (continuous blue line).}
\label{fig_igt_sparseDR}
\end{figure*}

We have compared the IHT and DR sparsity-based inpainting algorithms on 100 Monte-Carlo simulations using a mask with sky coverage $\fsky=77\%$. In all our experiments, we have used $150$ iterations for both iterative schemes, $\beta=1$ and $\alpha_n \equiv 1$ ($\forall n$) in the DR scheme (see Appendix B). For each inpainted map $i \in \{1,\cdots,100\}$, we computed the relative mean squared-error (MSE)
\begin{equation*}
e^{(i)}[\ell] =  \left<  \frac{  \abs{\cmba^{\mathrm{True}}_{\ell,m} - \cmba^{(i)}_{\ell,m} }^2 }{ C_{\ell} } \right>_m
\end{equation*}
and the its version in percent per $\ell$
\begin{equation*}
E[\ell] =    100 \times \left<   e^{(i)}[\ell]    \right>_i ~ (\%) ~.
\label{eq_err_inp}
\end{equation*}

Fig.~\ref{fig_igt_sparseDR} depicts the relative MSE in percent per $\ell$ for the two sparsity-based inpainting algorithms IHT ($l_0$) and DR ($l_1$). We see that at very low $\ell$, $l_1$-sparsity inpainting as provided by the DR algorithm yields better results. We have performed the test with other masks, and arrived at similar conclusions.
As this paper focuses on low-$\ell$, only the $l_1$ inpainting as solved by the DR algorithm will be considered in the sequel.

\section{Energy Prior}
\label{sec:energy}
\subsection*{Optimization problem}
If we know a priori the power-spectrum $(C_{\ell,m})_{\ell,m}$ (not the angular one, which implicitly assumes isotropy) of the Gaussian field $\cmb$, then using a maximum a posteriori (MAP) argument, the inpainting problem amounts to minimizing a weighted $\ell_2$-norm subject to a linear constraint
\begin{equation}
\label{eq:minwl2}
\text{find } \widehat{\cmb} = \argmin{\cmb \in \RR^n} \norm{\sht\cmb}_{C^{-1}} \st \obs = \mask \cmb ~,
\end{equation}
where for a complex-valued vector $\cmba$, $\norm{\cmba}^2_{C^{-1}} = \sum_{\ell,m} \frac{\abs{\cmba_{\ell,m}}^2}{C_{\ell,m}}$, i.e. a weighted $\ell_2$ norm. By strong convexity, Problem \eqref{eq:minwl2} is well-posed and has exactly one minimizer, therefore justifying equality instead of inclusion in the argument of the minimum in \eqref{eq:minwl2}.

Since the objective is differentiable with a Lipschitz-continuous gradient, and the projector $\proj_{\{\cmb: \obs = \mask \cmb\}}$ on the linear equality set is known in closed-form, one can use a projected gradient-scheme to solve \eqref{eq:minwl2}. However, it turns that the estimate of the descent step-size can be rather crude, which may hinder the convergence speed. One can think of using an inexact line-search but this will make the algorithm unnecessarily complicated. 
This is why we propose in Appendix C a new algorithm, based  on the Douglas-Rachford splitting method,  which is easy to implement and efficient.

In all our experiments in the rest of the paper, we have used $150$ iterations for this algorithm with $\beta=50$ and $\alpha_n \equiv 1$ ($\forall n$) (see Appendix C). 
This approach requires the power-spectrum $C_{\ell,m}$ as an input parameter. In practice, an estimate $\tilde{C}_{\ell}$ of the power spectrum from the data using MASTER was used, and we set $C_{\ell,m} =  \tilde{C}_{\ell}$. The latter is reminiscent of an isotropy assumption which is not necessarily true.

\section{Isotropy Prior}
\label{sec:isotropy}
\subsection{Structural constraints on the power spectrum}
Strictly speaking, we would define the set of isotropic homogeneous random processes on the (discrete) sphere as the closed set
\begin{equation*}
\label{eq:Ciso}
\cC_{\mathrm{iso}} := \ens{\cmb: C_{\ell,m} = C_{\ell}, ~ \forall -\ell \leq m \leq \ell, } .
\end{equation*}
where $C_{\ell,m}= \E{\abs{(\sht\cmb)_{\ell,m}}^2}$ and $C_{\ell}$ is the angular power spectrum, which depends solely on $\ell$. Given a realization $\cmb$ of a random field, we can then naively state the orthogonal projection of $\cmb$ onto $\cC_{\mathrm{iso}}$ by solving
\begin{equation*}
\min_{\boldsymbol{v} \in \cC_{\mathrm{iso}}} \norm{\boldsymbol{v}-\cmb}_2 ~.
\end{equation*}
This formulation of the projection constraint set is not straightforward to deal with for at least the following reasons: (i) the isotropy constraint set involves the unknown true $C_\ell$ (through the expectation operator), which necessitates to resort to stochastic programming; (ii) the constraint set is not convex. 

\subsection{Projection with a deterministic constraint}
An alternative to the constraint set $\cC_{\mathrm{iso}}$ would be to replace $C_{\ell,m}$ and $C_{\ell}$ by their empirical sample estimates, i.e. $C_{\ell,m}$ by $\abs{\cmba_{\ell,m}}^2$ and $C_{\ell}$ by $\widehat{C}_{\ell} = \frac{1}{2\ell+1}\sum_{m}\left|\cmba_{\ell,m}\right|^{2}$. However, this hard constraint might be too strict in practice and we propose to make it softer by taking into account variability inherent to the sample estimates, as we explain now.

\medskip

In a nutshell, the goal is to formulate a constraint set, where for each $\ell$, the entries $\cmba_{\ell,m}$ of the spherical harmonic coefficient vector $\cmba$ deviate the least possible in magnitude (up to a certain accuracy) from some pre-specified vector $\vmu$; typically we take $\vmu = \sqrt{C_{\ell}}$ or its empirical estimate $\sqrt{\widehat{C}_{\ell}}$. Put formally, this reads
\begin{equation*}
\label{eq:Cisoemp}
\cC_\veps = \ens{\cmb: \cmba=\sht \cmb, \abs{\abs{\cmba_{\ell,m}} - \vmu_{\ell}} \leq \veps, ~ \forall \ell, m}  ~.
\end{equation*}
$\cC_\veps$ is a compact set, although not convex. \\

\begin{figure}[htb]
\centering
\includegraphics[width=0.4\textwidth]{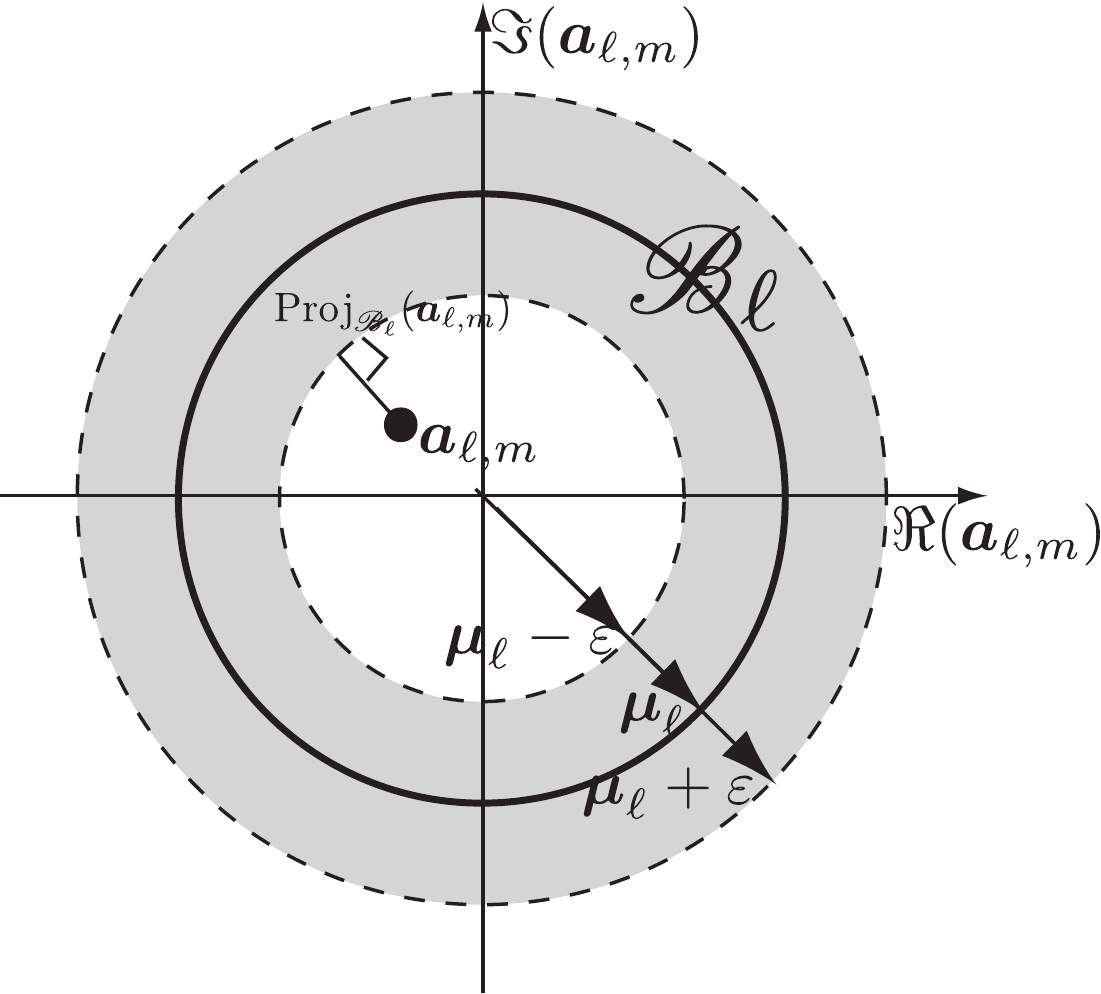}
\caption{The set $\mathscr{B}_{\ell}$.}
\label{fig:projectorband}
\end{figure}

We now turn to the projection on $\cC_\veps$. We begin by noting that this set is separable, i.e. 
$\mathcal{C}_{\veps} = {\times}_{\ell,m} \mathscr{B}_{\ell}$, 

where $\mathscr{B}_{\ell}$ is the band depicted Fig.~\ref{fig:projectorband}. 
It turns out that for fixed $\vmu$, $\mathscr{B}_{\ell}$ is so-called prox-regular since its associated orthogonal projector is single-valued with a closed-form. Indeed, the projector onto $\mathcal{C}_{\veps}$ is
\begin{equation}
\proj_{\mathcal{C}_{\veps}}(\cmb) = \sht^*\parenth{ \proj_{\mathscr{B}_{\ell}}\parenth{\cmba_{\ell,m}}}_{\ell,m},
\end{equation}
where, 
\be
\label{eq:projBl}
\proj_{\mathscr{B}_{\ell}}(\cmba_{\ell,m}) & = &   
 \begin{cases}
(\vmu_{\ell} + \veps)\cmba_{\ell,m}/\abs{\cmba_{\ell,m}} & \\
 ~~~~ \text{if} ~ \abs{\cmba_{\ell,m}} > \vmu_{\ell} + \veps &   \\
(\vmu_{\ell} - \veps)\cmba_{\ell,m}/\abs{\cmba_{\ell,m}} &  \\
~~~~ \text{if} ~ \abs{\cmba_{\ell,m}} < \vmu_{\ell} - \veps &   \\
\cmba_{\ell,m} ~~ \text{otherwise.}&
\end{cases}
\ee

\paragraph{\bf Choice of the constraint radius:}
To devise a meaningful choice (from a statistical perspective) of the constraint radius $\veps$, we first need to derive the null distribution\footnote{That is the distribution under the isotropy hypothesis.} of $\abs{\cmba_{\ell,m}} - \vmu_{\ell}, \forall (\ell,m)$, with the particular case when $\vmu_{\ell}=\sqrt{\widehat{C}_{\ell}}$. $\cmba_{\ell,m}$ being the spherical harmonic coefficients of a zero-mean stationary and isotropic Gaussian process whose angular power spectrum is $C_{\ell}$, it is easy to show that
\begin{multline*}
2\abs{\cmba_{\ell,m}}^2  \sim C_{\ell}\chi^2(2)   \quad \text{and} \quad \\ 
2L\vmu_{\ell}^2        \sim C_{\ell}\chi^2(2L), ~ \forall \ell \geq 2 ~,
\end{multline*}
where $L=2\ell+1$.
By the Fisher asymptotic formula, we then obtain
\begin{eqnarray*}
\abs{\cmba_{\ell,m}} & \overset{d}{\rightarrow}&  \sqrt{C_{\ell}}\GN\parenth{\sqrt{\tfrac{3}{4}},\tfrac{1}{4}}    \quad \text{and} \quad  \nonumber \\ 
\vmu_{\ell}         &  \overset{d}{\rightarrow} & \sqrt{C_{\ell}}\GN\parenth{\sqrt{1-\tfrac{1}{4L}}, \tfrac{1}{4L}}, \quad \forall \ell \geq 2 ~,  
\end{eqnarray*}
where $\overset{d}{\rightarrow}$ means convergence in distribution. Thus, ignoring the obvious dependency between $\abs{\cmba_{\ell,m}}$ and $\vmu_{\ell}$, we get
\begin{eqnarray*}
\abs{\cmba_{\ell,m}} - \vmu_{\ell} \overset{d}{\rightarrow} \sqrt{C_{\ell}}\GN\parenth{\sqrt{\tfrac{3}{4}}-    \sqrt{1- \tfrac{1}{4L}}, \tfrac{L+1}{4L}}~.
\end{eqnarray*}
Note that the distribution of this difference can be derived properly using the delta method in law, but computing the covariance matrix of the augmented vector $\parenth{\cmba_{\ell,m}}_m$, remains an issue. The quality of the above asymptotic approximation becomes better as $\ell$ increases.

The upper and lower critical thresholds at the double-sided significance level $0 \leq \alpha \leq 1$ are 
\begin{multline}
\veps_{\pm} = \sqrt{C_{\ell}}  \bigg(\sqrt{\tfrac{3}{4}}-  \sqrt{1- \tfrac{1}{4L}} \pm \sqrt{  \tfrac{L + 1}{4L}}\Phi^{-1}(1-\tfrac{\alpha}{2})\bigg) \nonumber
\end{multline}
where $\Phi$ is the standard normal cumulative distribution function. 

\begin{figure*}
\centering
\includegraphics[width=0.75\textwidth]{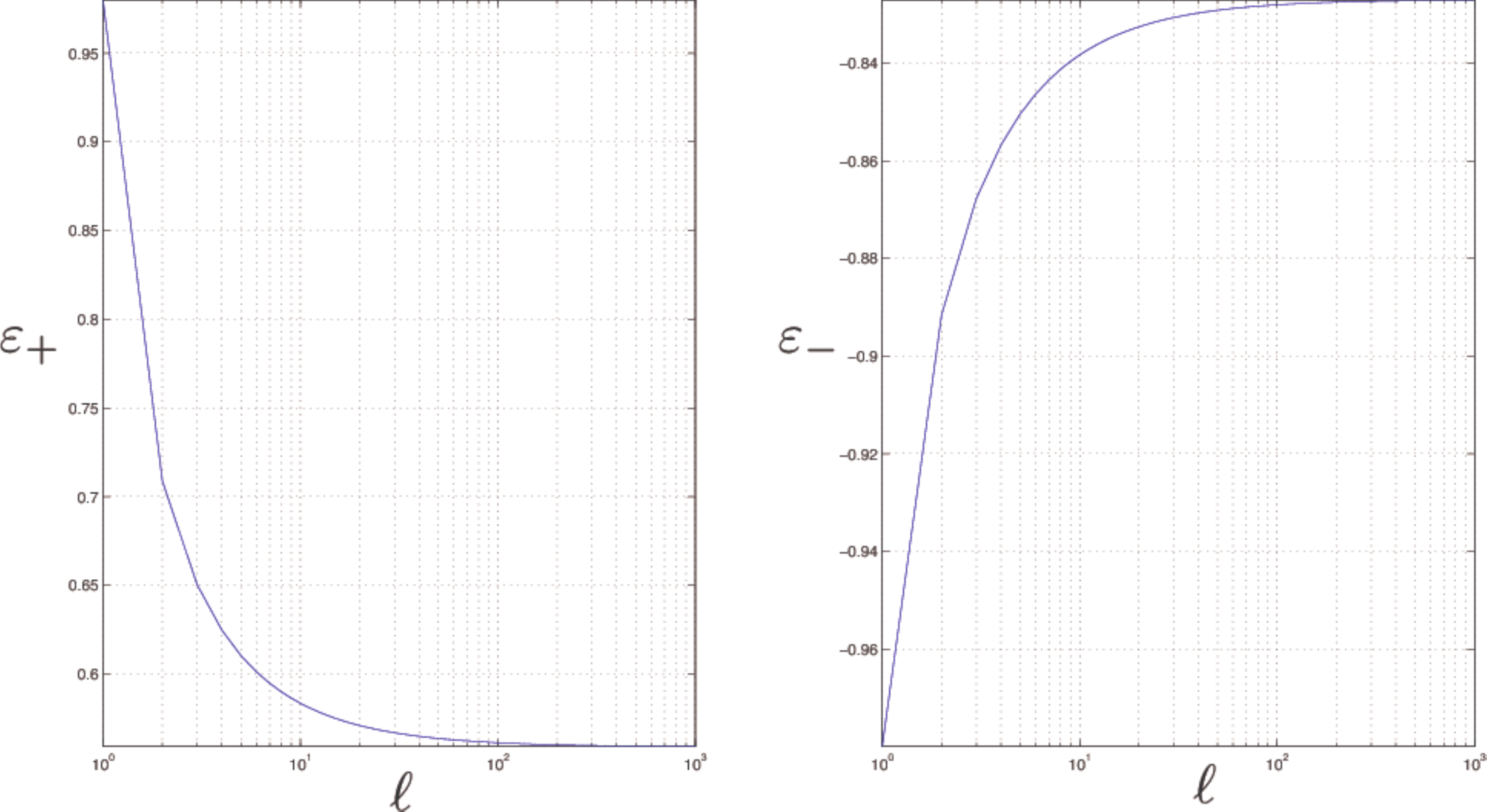}
\caption{The upper (left) and lower (right) normalized critical thresholds at the significance level $0.05$ as a function of $\ell$.}
\label{figalpha0p05}
\end{figure*}

We depict in Fig~\ref{figalpha0p05} the upper and lower critical thresholds normalized by $\sqrt{C_{\ell}}$ at the classical significance level $0.05$ as a function of $\ell$. One can observe that the thresholds are not symmetric. This entails that different values of $\veps$ should be used in \eqref{eq:projBl}. Furthermore, as expected, the two thresholds decrease in magnitude as $\ell$ increases. They attain a plateau for $\ell$ large enough (typically $\geq 100$). It is easy to see that the two limit values are $\sqrt{3/4}-1+1.96/\sqrt{8}=0.559$ and $\sqrt{3/4}-1-1.96/\sqrt{8}=-0.8269$. 

The isotropy-constrained inpainting algorithm is given is Appendix D.
 

\section{Experiments}

\subsection{$a_{\ell,m}$  Reconstruction}

\begin{figure*}[htb]
\vbox{
\hbox{
\includegraphics [scale=0.3]{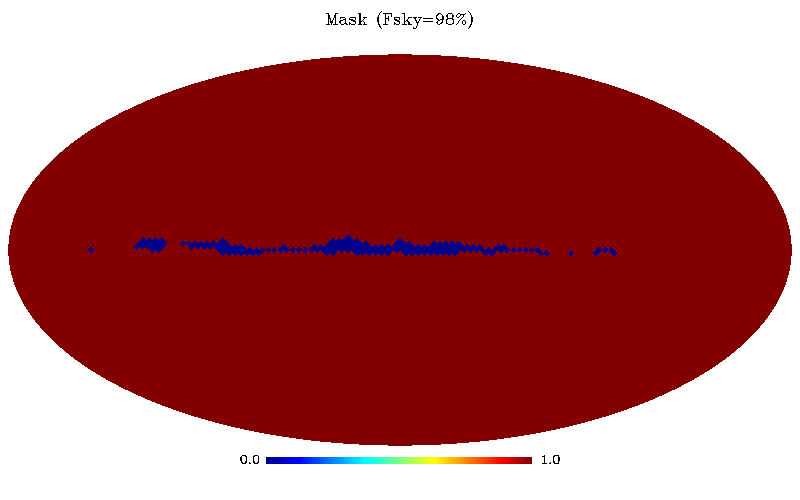}
\includegraphics [scale=0.3]{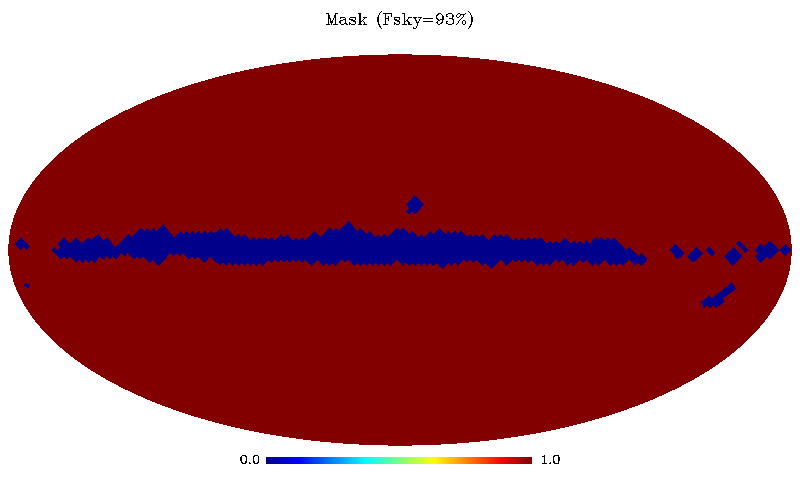}}
\hbox{
\includegraphics [scale=0.3]{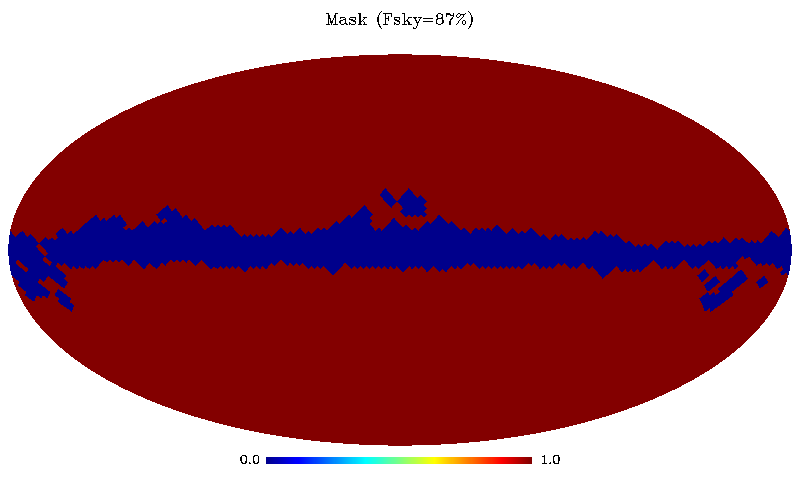}
\includegraphics [scale=0.3]{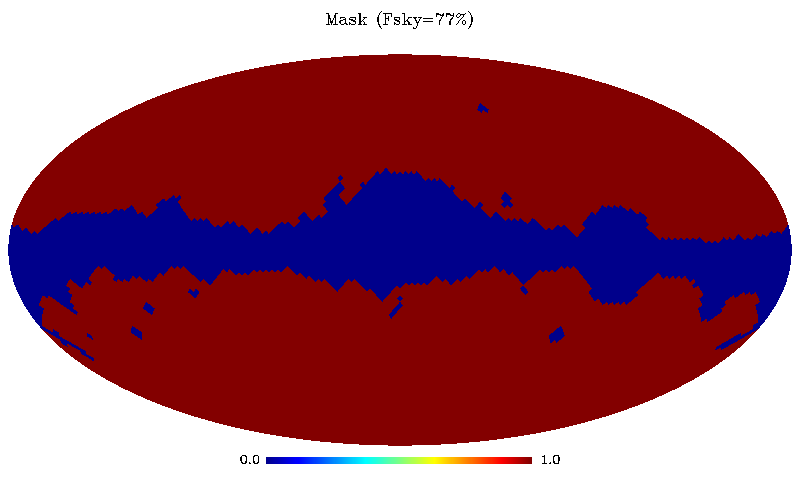}}
\hbox{
\includegraphics [scale=0.3]{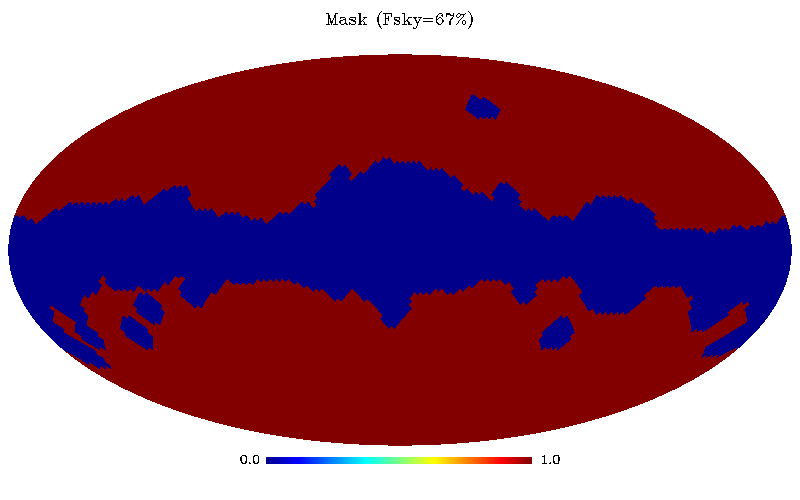}
\includegraphics [scale=0.3]{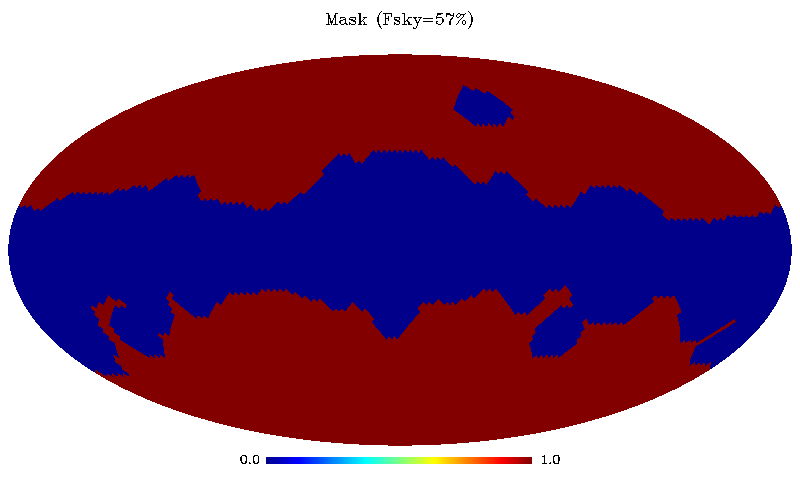}}
}
\caption{Six masks with respectively $\fsky \in \{0.98,0.93,0.87,0.77,0.67,0.57\}$. The masked area is in blue.}
\label{fig_mask}
\end{figure*}

\begin{figure*}[htb]
\vbox{
\hbox{
\includegraphics [scale=0.3]{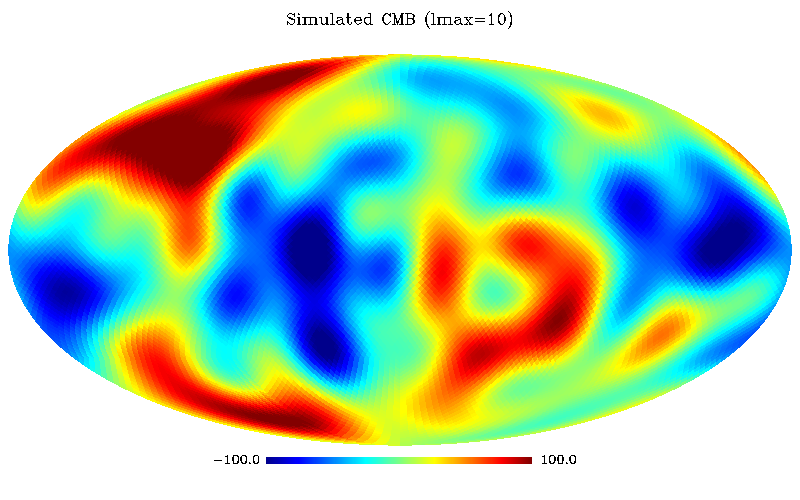}
\includegraphics[scale=0.3]{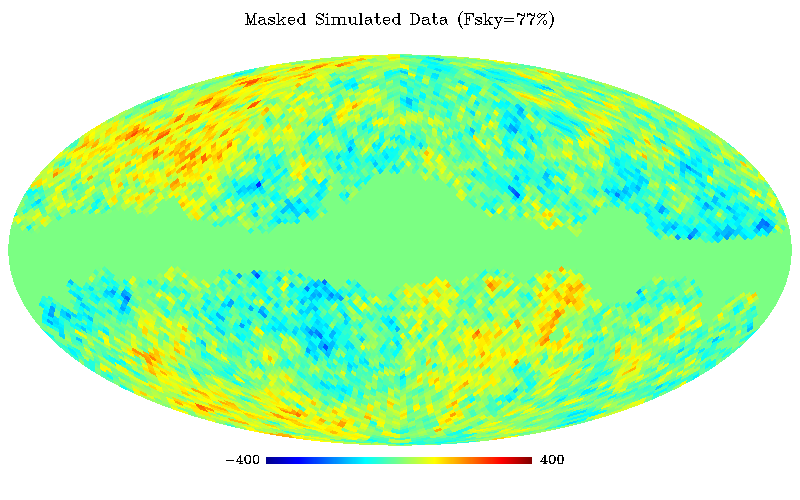}}
\hbox{
\includegraphics [scale=0.3]{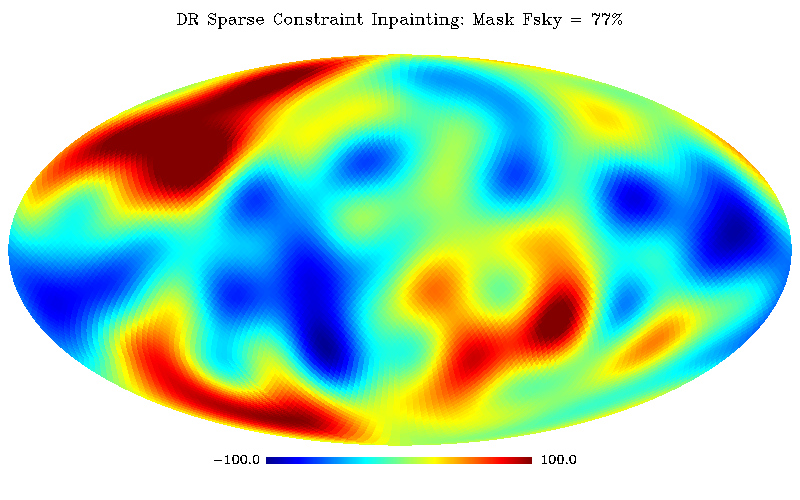}
\includegraphics [scale=0.3]{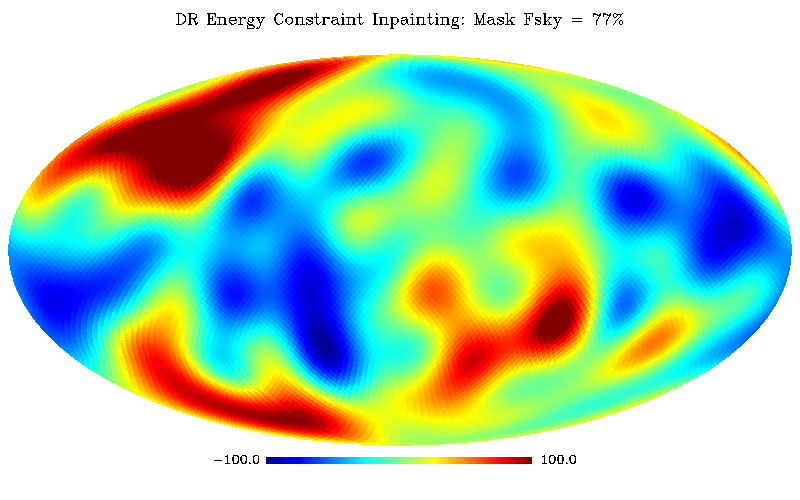}}
\hbox{
\includegraphics [scale=0.3]{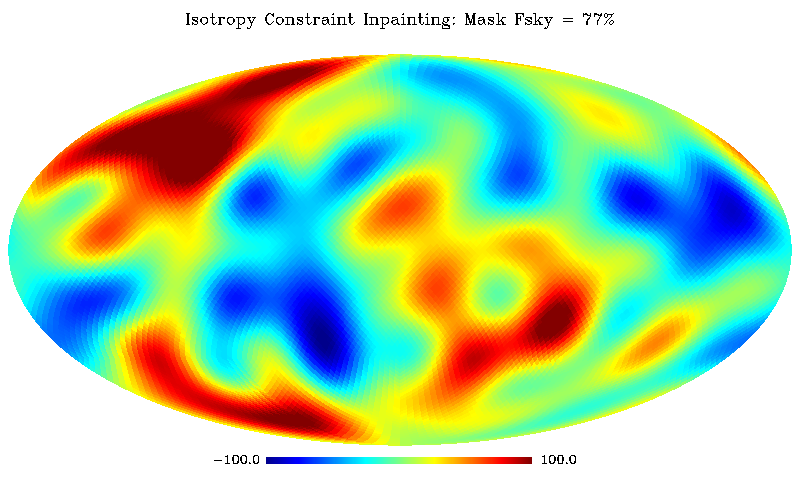}
\includegraphics [scale=0.3]{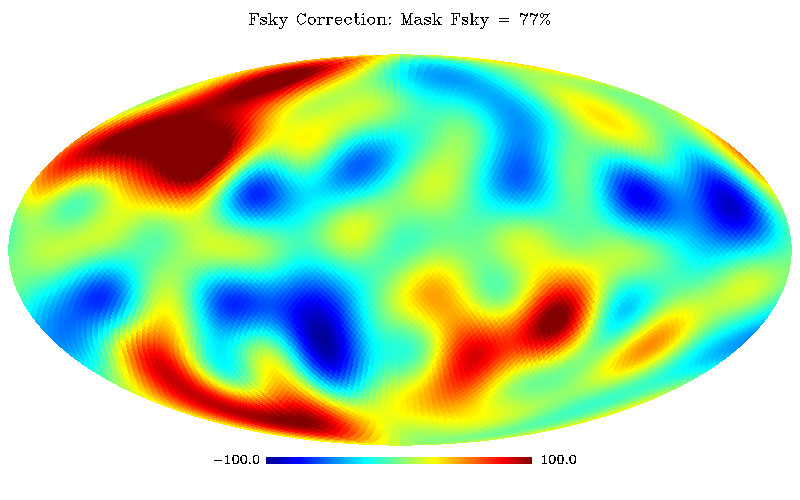}}
}
\caption{Top, input smoothed simulated CMB map  ($\ell_{\max}=10$) and the same map, but masked ($\fsky=77$\%) and not smoothed (i.e. input simulated data). Middle, inpainting of the top right image up to $\ell_{\max}=10$, using a sparsity prior (left) and an energy prior (right). Bottom, inpainting of the top right image up to $\ell_{\max}=10$, using an isotropy prior (left) and a simple $\fsky$ correction (right). }
\label{fig_inp_mask_4methods}
\end{figure*}

\begin{figure*}[htb]
\vbox{
\hbox{
\includegraphics [scale=0.3]{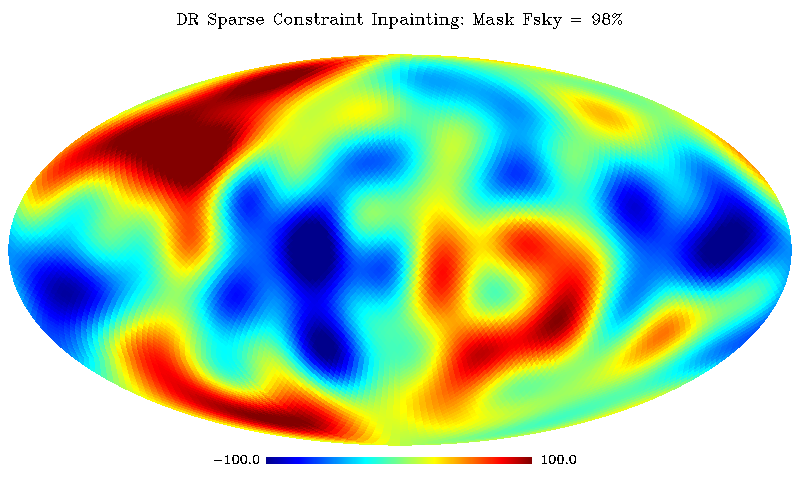}
\includegraphics [scale=0.3]{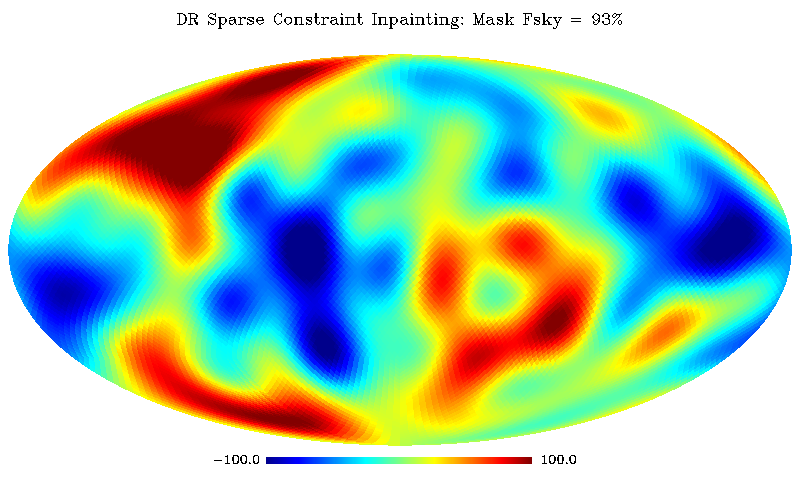}}
\hbox{
\includegraphics[scale=0.3]{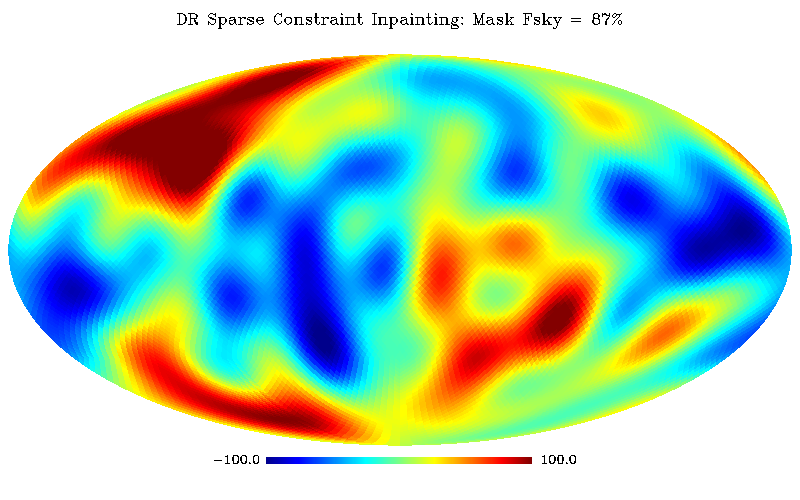}
\includegraphics [scale=0.3]{./images/fig_dr_sparse_fsky_77.png}}
\hbox{
\includegraphics[scale=0.3]{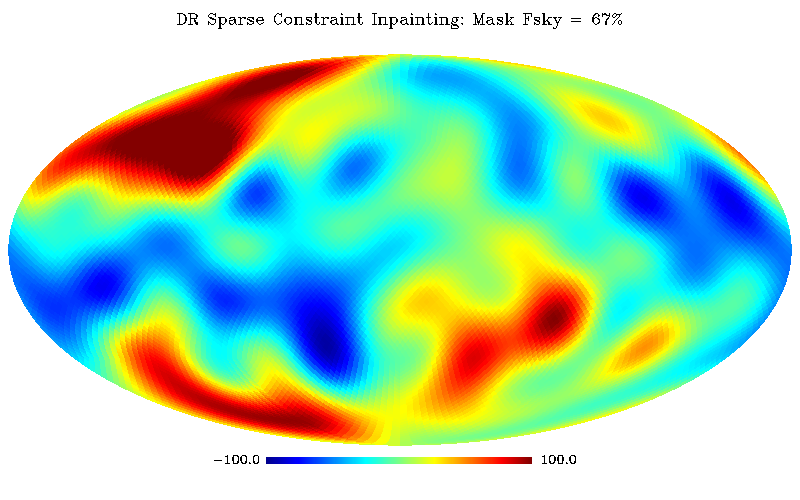}
\includegraphics [scale=0.3]{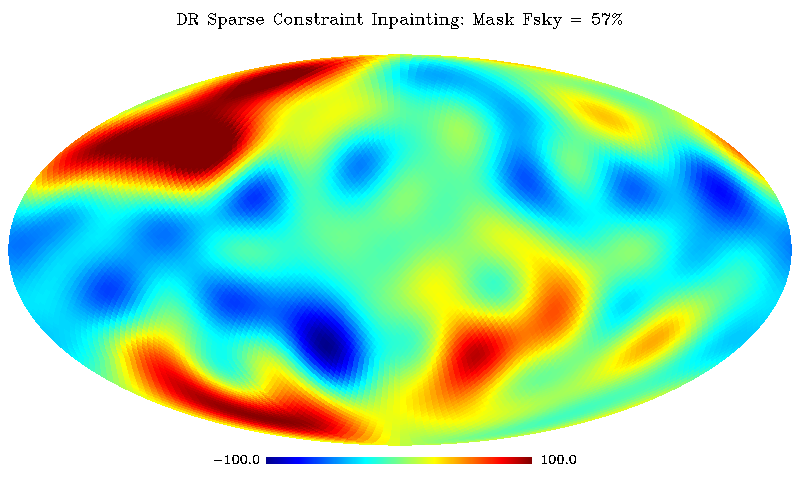}}
}
\caption{$\ell_1$ sparsity-based inpainting of a simulated CMB map with different masks. Top, $\fsky$ = 98\% and 93\%, middle 87\% and 77\%, and bottom 67\% and 57\%.}
\label{fig_sparse_inp_6mask}
\end{figure*}

In this section, we compare the different inpainting methods described previously: the DR-sparsity prior inpainting, the DR-energy prior inpainting and the isotropy prior inpainting. 
We also test the case of no-inpainting, applying just an $\fsky$ correction to the $\cmba_{\ell,m}$ spherical harmonic coefficients of the map
(i.e.  data $a_{\ell,m}$ are corrected with a  correction term equal to $ 1 / \sqrt(\mathrm{Fsky}$).
We have run 100 CMB Monte-Carlo simulations with a resolution corresponding to nside equal to 32, which is precise enough for the large scales we are considering. We have considered six different masks, with sky coverage $\fsky$ ranging in $\{0.98,0.93,0.87,0.77,0.67,0.57\}$.

Fig~\ref{fig_mask} displays these masks. Note that point source masks have not been considered here, since they should not affect $\cmba_{\ell,m}$ estimation at very low multipole. Each of these simulated CMB maps has been masked with the six masks.  Fig.~\ref{fig_inp_mask_4methods} top right shows one CMB realization masked with the $77$\% $\fsky$ mask.
Fig.~\ref{fig_inp_mask_4methods} middle and bottom show respectively the inpainting of the Fig.~\ref{fig_inp_mask_4methods} top right image, with the four methods (i.e. using sparsity, energy, isotropy priors, and $\fsky$ correction). Fig.~\ref{fig_sparse_inp_6mask} depicts the results given by the $\ell_1$ sparsity-based inpainting method for the six different masks. We can clearly see how the quality of the reconstruction degrades with decreasing sky coverage.

\begin{figure*}[htb]
\vbox{
\hbox{
\includegraphics [trim= 1.5cm 13cm 2cm  2.6cm , clip,  scale=0.4]{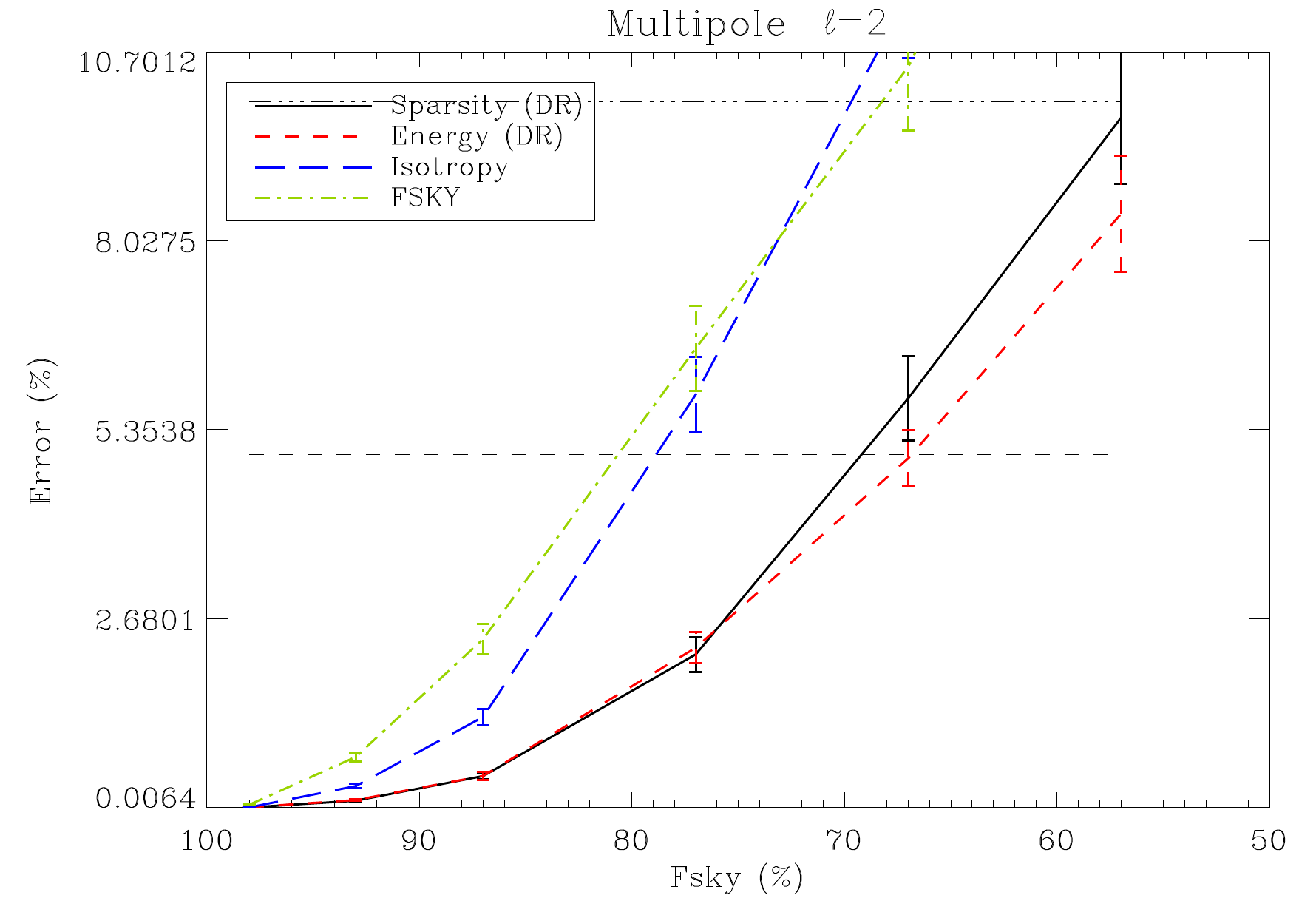}
\includegraphics [trim= 1.5cm 13cm 2cm  2.6cm , clip,  scale=0.4]{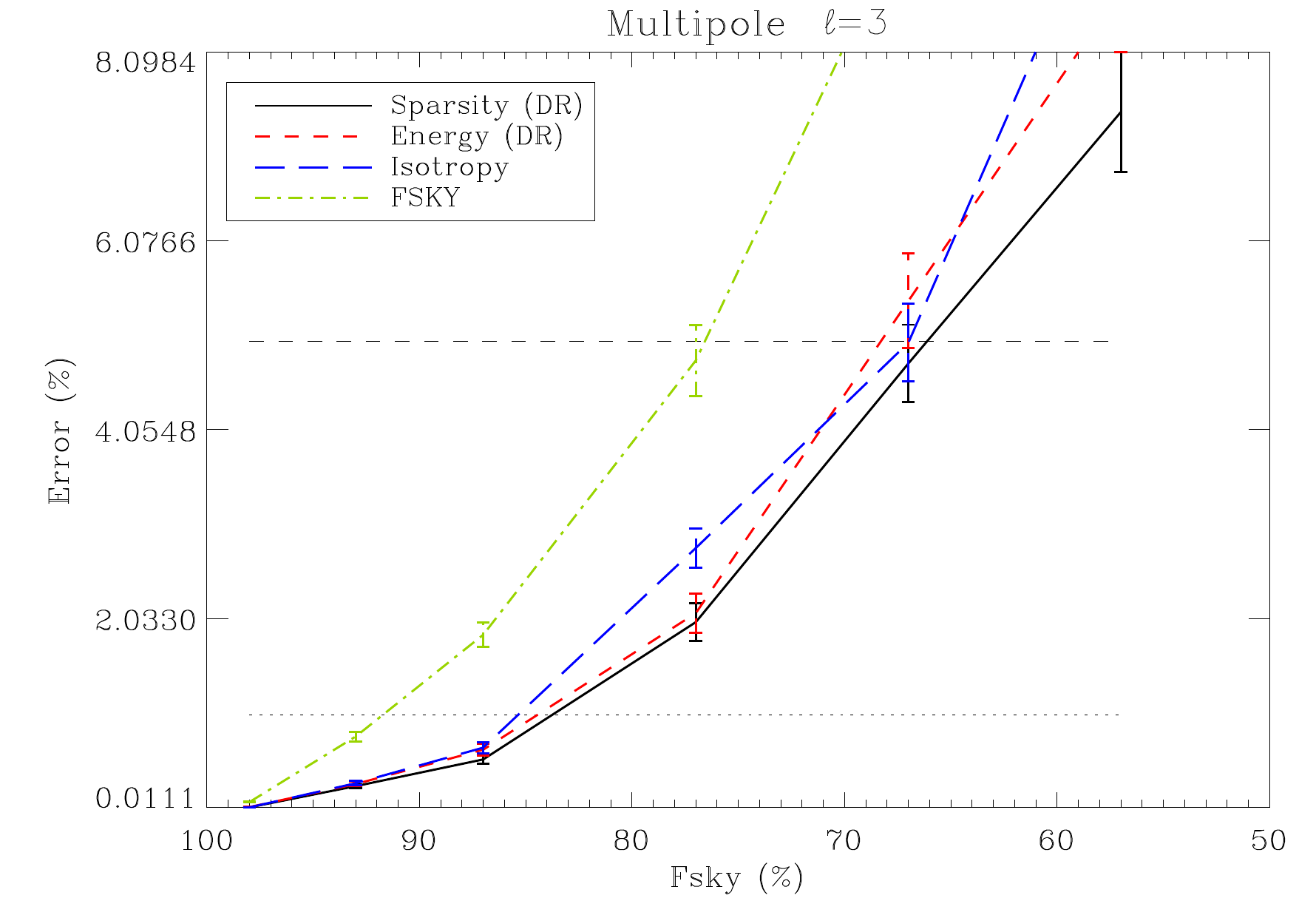}}
\hbox{
\includegraphics [trim= 1.5cm 13cm 2cm  2.6cm , clip,  scale=0.4]{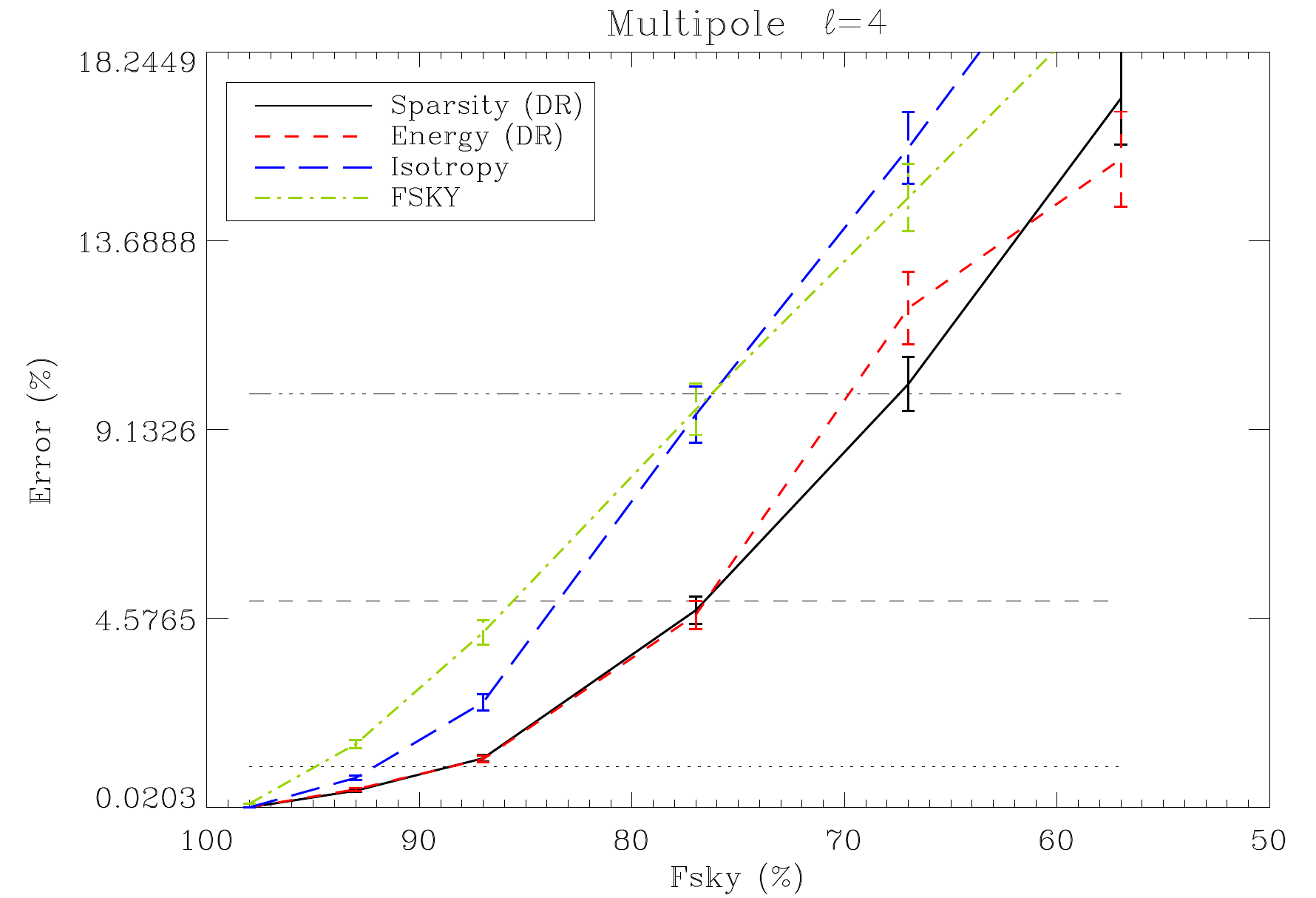}
\includegraphics [trim= 1.5cm 13cm 2cm  2.6cm , clip,  scale=0.4]{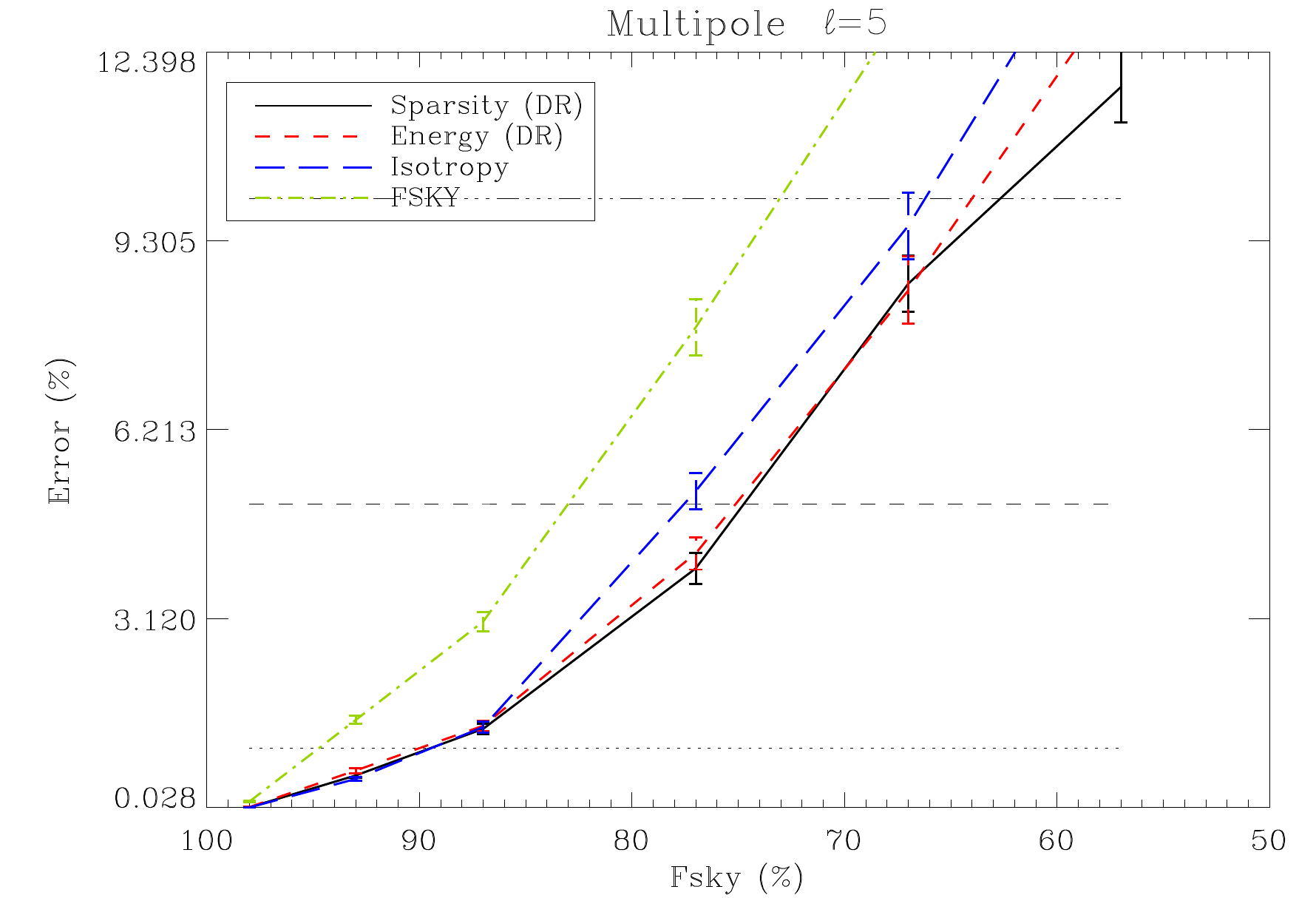}}
}
\caption{Relative MSE (in percent) per $\ell$ versus sky coverage for different multipoles. The three horizontal lines correspond to 1\%, 5\% and 10\%  of the cosmic variance. Reconstruction within the 1\% of the cosmic variance can be achieved up to $\ell=4$ using the inpainting methods based on the sparsity or the energy priors  for a mask with $\fsky$ larger than $80\%$. With a 77\% coverage Galactic mask, the error increases by a factor 5 at $\ell=4$.}
\label{fig_err_inp_vers_fsky}
\end{figure*}

The $\cmba_{\ell,m}$ coefficients of the six hundred maps (100 realizations masked with the six different masks) 
have then been estimated using the three inpainting methods and the $\fsky$ correction methods. 
The approaches were compared in terms of the relative MSE and relative MSE per $\ell$; see end of Section~\ref{sec:sparsity} for their definition. 

Fig.~\ref{fig_err_inp_vers_fsky} shows relative MSE for $\ell=2$ to 5 versus the sky coverage $\fsky$. 
The three horizontal lines correspond to 1\%, 5\% and 10\%  of the cosmic variance (CV). 
Inpainting based on the sparsity and energy priors give relatively close results, and anyway better than the one assuming the isotropy prior or the $\fsky$ correction.
 It is interesting to notice that a very high quality reconstruction (with 1\% of the CV) can be obtained with both the sparsity- and energy-based inpainting methods 
 up to $\ell=4$, for a mask with $\fsky$ larger than 80\%. 
 If WMAP data do not allow us to make non-Gaussianity studies with such a small mask, Planck component separation will certainly be able to achieve the required quality 
 to make possible the use of such a small Galactic mask. With a 77\% coverage Galactic mask, the error increases by a factor 5 at $\ell=4$ !

\begin{figure*}[htb]
\vbox{
\hbox{
\includegraphics [trim= 3cm 13cm 2cm  2.6cm , clip,  scale=0.4]{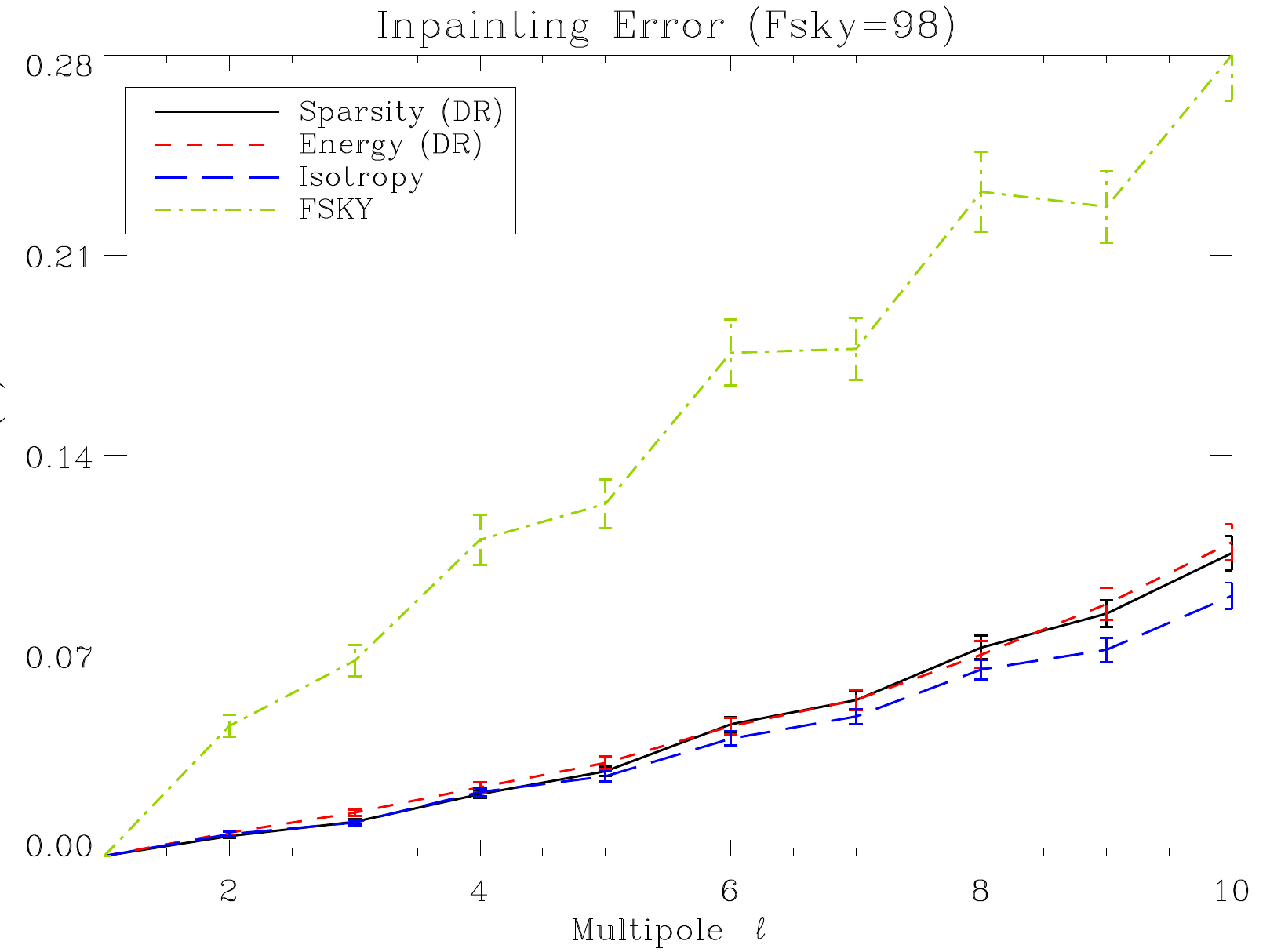}
\includegraphics [trim= 3cm 13cm 2cm  2.6cm , clip,  scale=0.4]{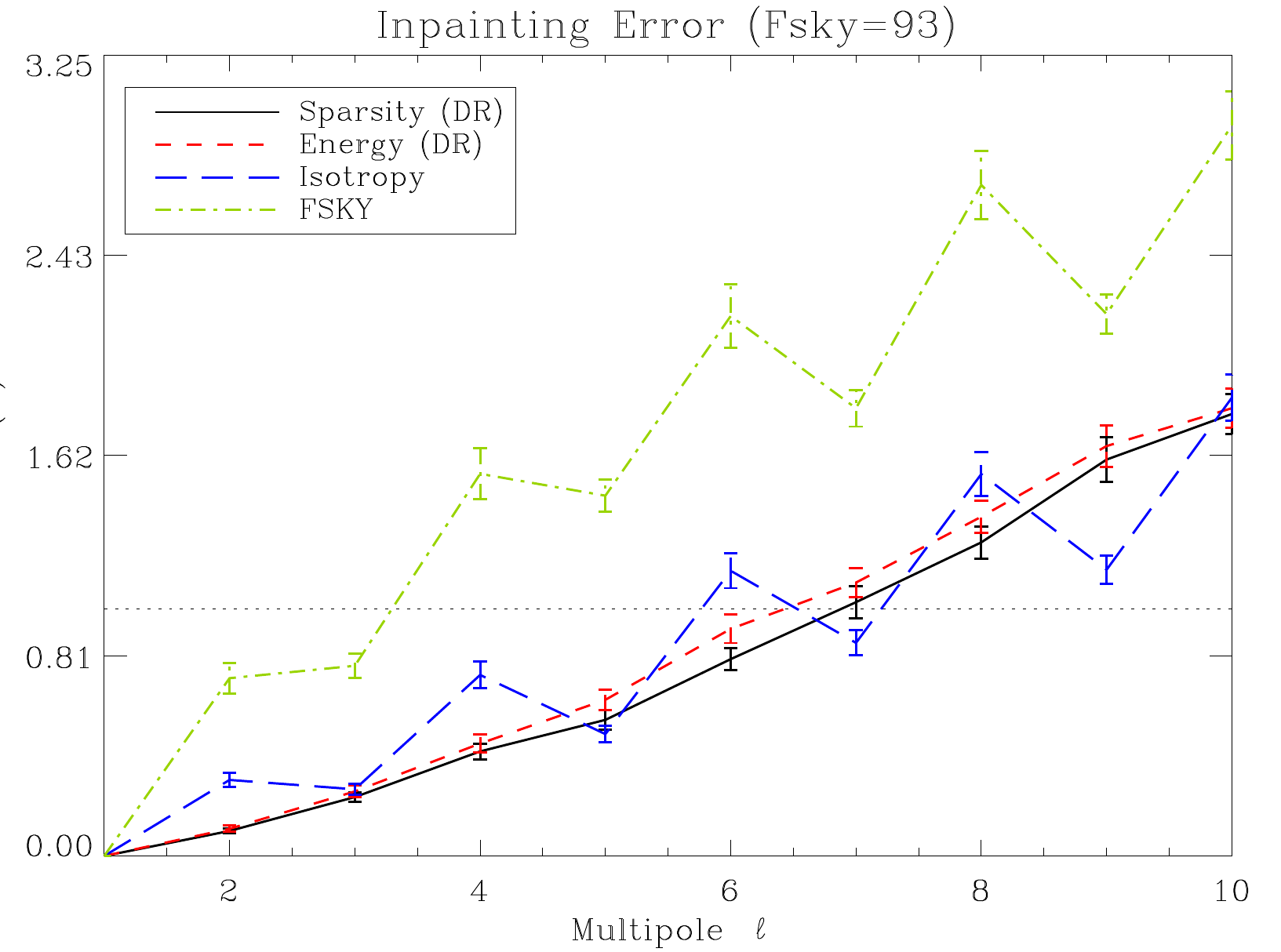}}
\hbox{
\includegraphics [trim= 3cm 13cm 2cm  2.6cm , clip,  scale=0.4]{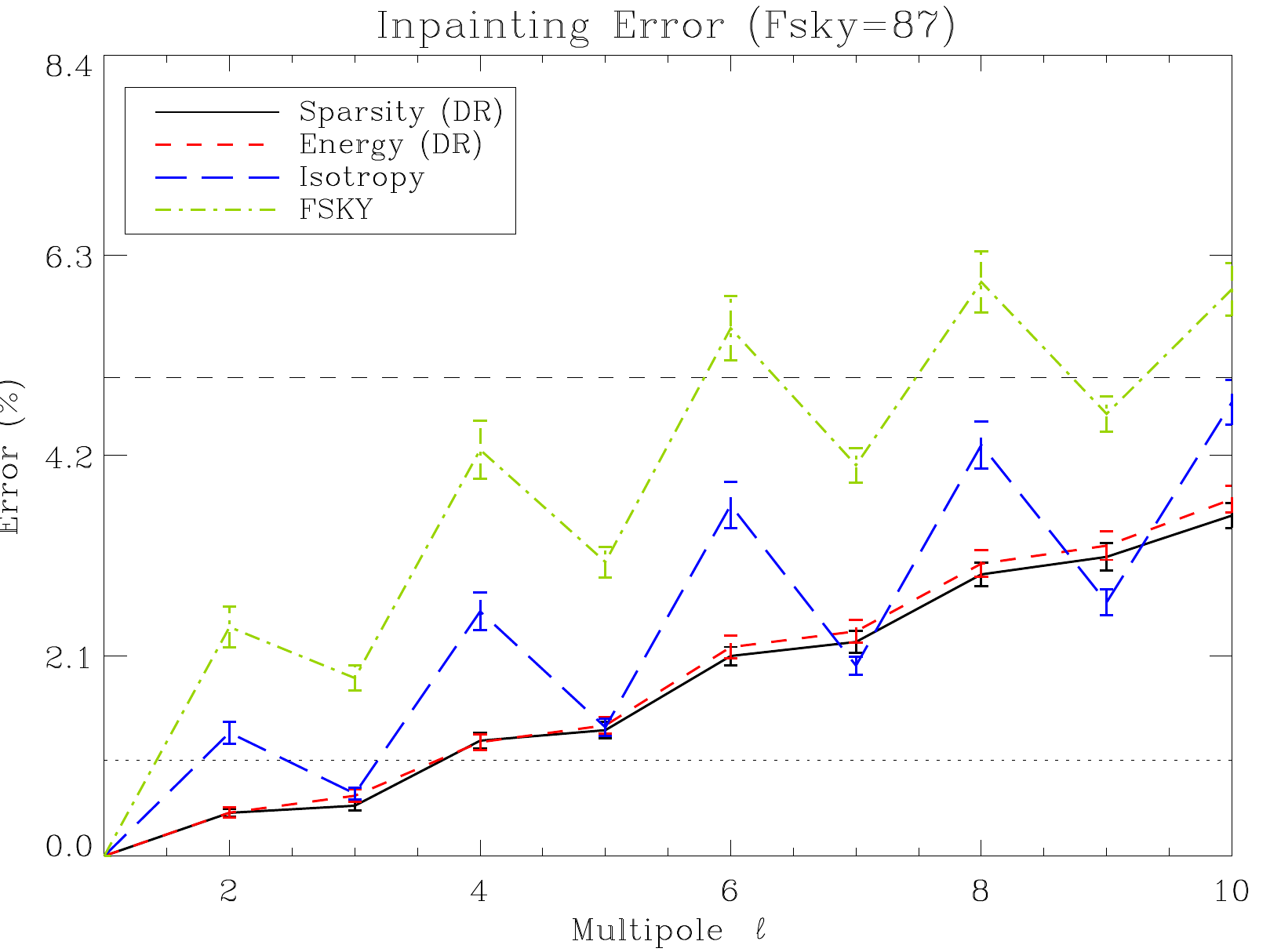}
\includegraphics [trim= 3cm 13cm 2cm  2.6cm , clip,  scale=0.4]{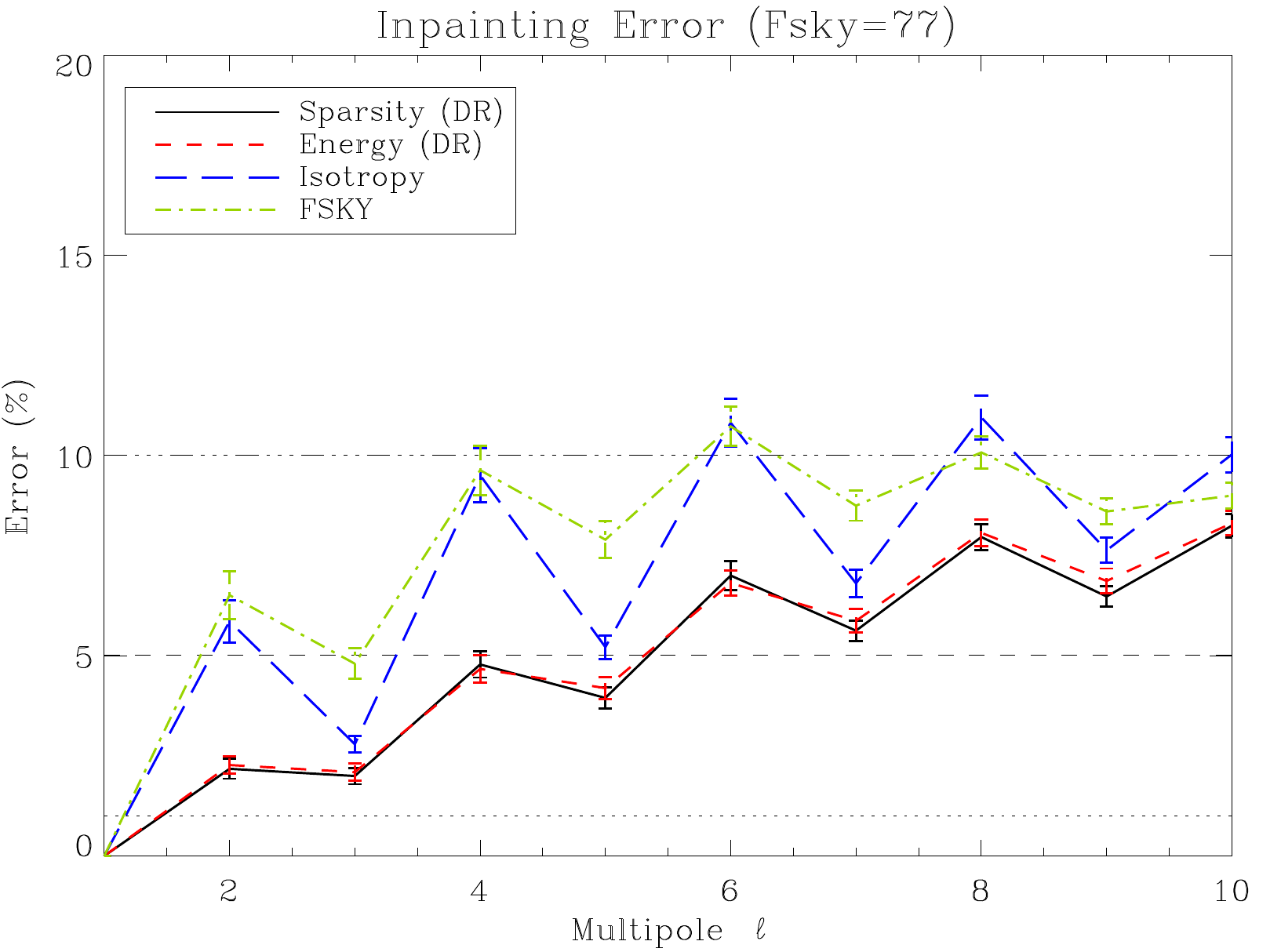}}
\hbox{
\includegraphics [trim= 3cm 13cm 2cm  2.6cm , clip,  scale=0.4]{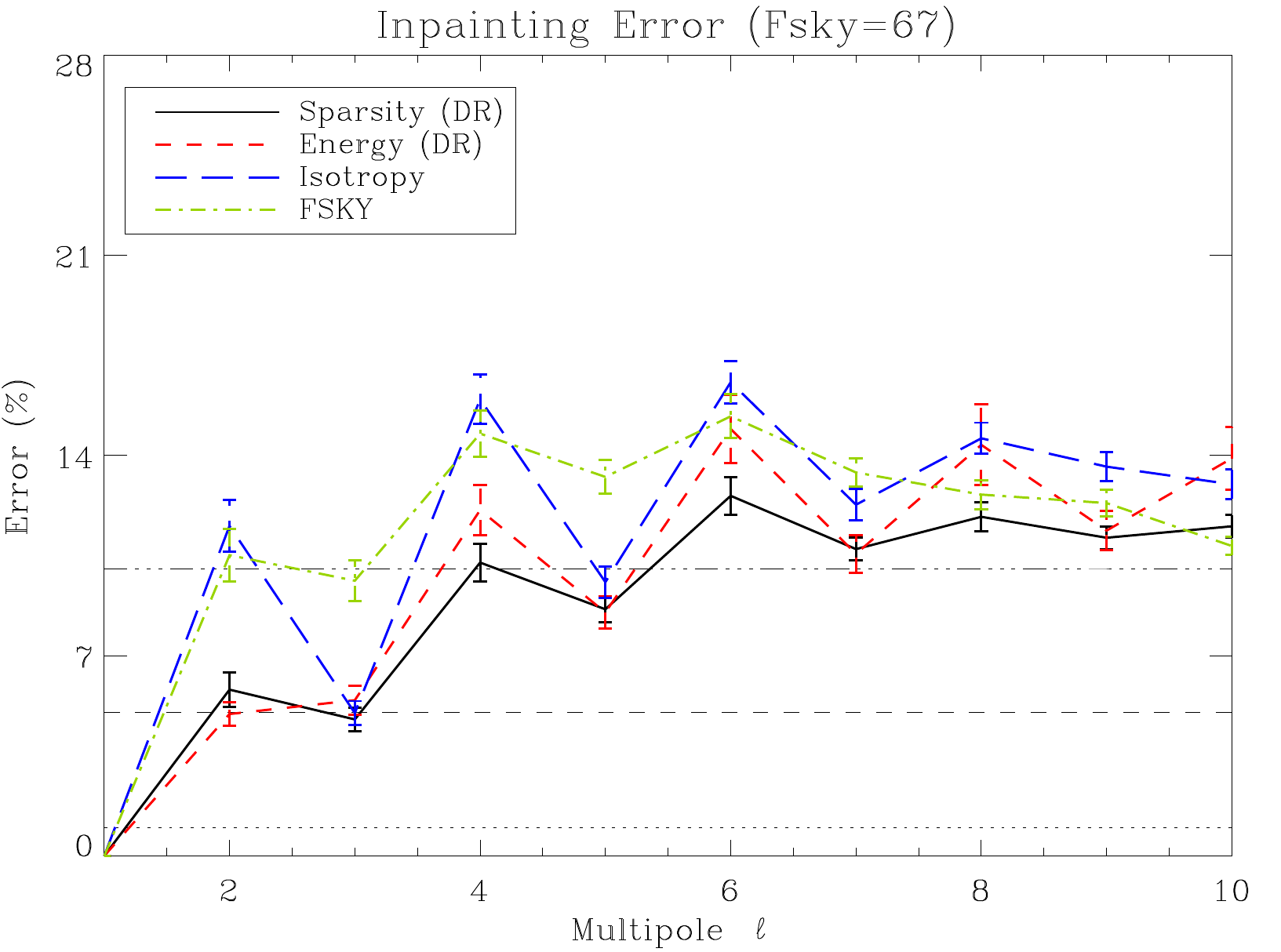}
\includegraphics [trim= 3cm 13cm 2cm  2.6cm , clip,  scale=0.4]{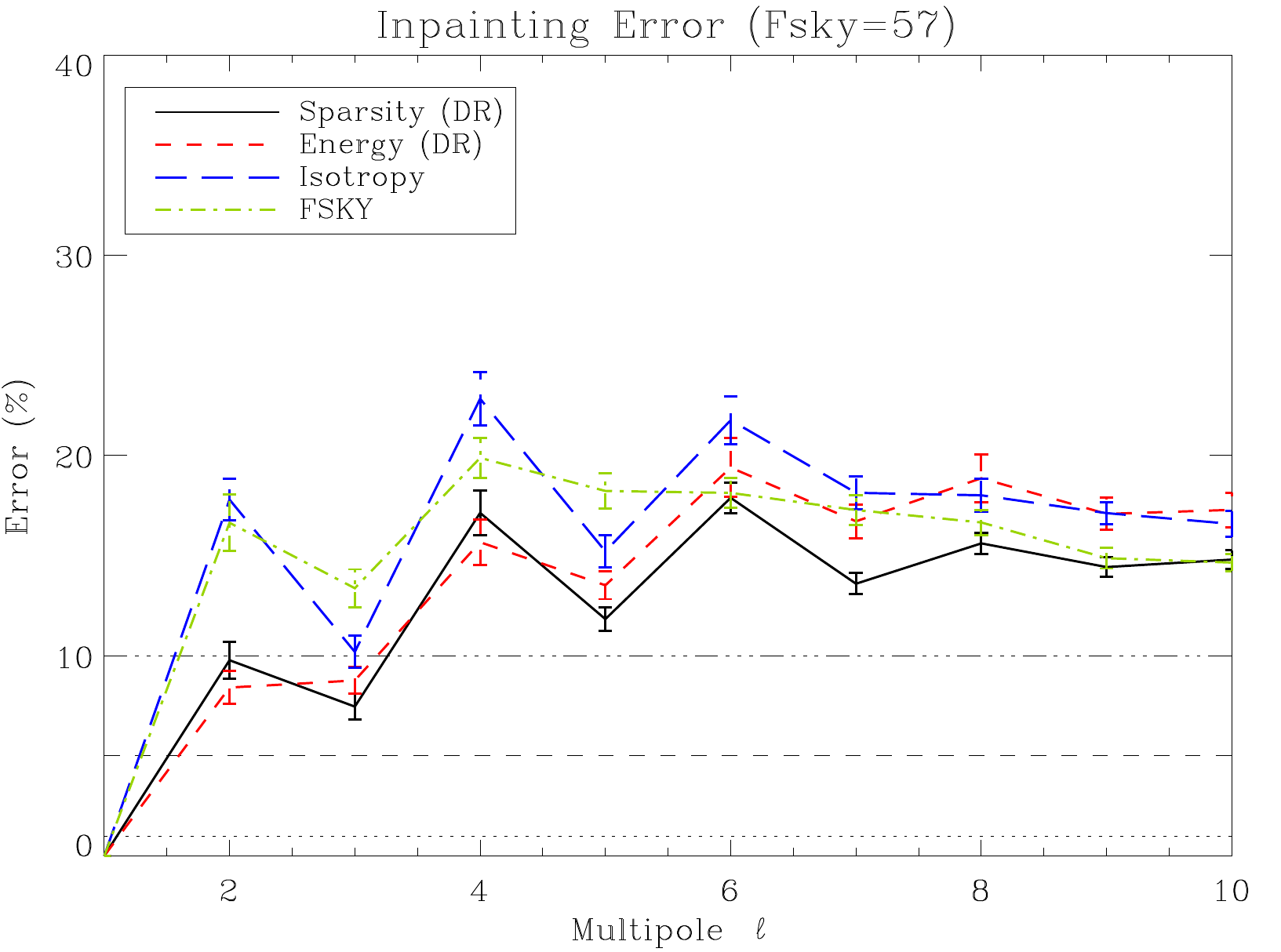}}
}
\caption{Relative MSE in percent per $\ell$ versus $\ell$ for the six masks.}
\label{fig_err_inp_vers_l}
\end{figure*}

Fig.~\ref{fig_err_inp_vers_l} shows the same errors, except that they are now plotted versus the multipoles for the six
different masks. The $\ell_1$ sparsity-based inpainting method seems to be slightly better than the one based on the energy prior, especially when the sky coverage decreases.

\subsection{Anomalies in CMB maps}

One the main motivations {of recovering all sky maps is to be able to study statistical properties such as Gaussianity and Statistical Isotropy on large scales. Statistical Isotropy is violated if there exists a preferred axis in the map. Mirror parity,  i.e. parity with respect to reflections through a plane: $\hat{\bf x}=\hat{\bf x} - 2(\hat{\bf x}\cdot\hat{\bf n})\hat{\bf n}$, where $\hat{\bf n}$ is the normal vector to the plane, is an example of a statistic where a preferred axis can be sought \citep{Land:2005odd,david2012}. 
 }
 
  With all-sky data, one can estimate the $S$-map for a given multipole in spherical harmonics, by considering \citep{david2012}: 
\begin{equation} \tilde{S}_\ell(\hat{\bf n}) = \sum_{m=-\ell}^{\ell}(-1)^{\ell+m}\frac{|a_{\ell m}(\hat{\bf n})|^2}{{\hat C}_\ell},\label{eq:paritymap}\end{equation}
where $a_{\ell m}(\hat{\bf n})$ corresponds to the value of the $a_{\ell m}$ coefficients when the map is rotated to have $\hat{\bf n}$ as the $z$-axis.
Positive (negative) values of $\tilde{S}_\ell (\hat{\bf n})$ correspond to even (odd) mirror parities in the $\hat{\bf n}$ direction. 
The same statistic can also be considered summed over all low multipoles one wishes to consider { \cite[e.g. focussing only on low multipoles as in][]{david2012}}: 
\begin{equation} 
\tilde{S}_{\rm tot}(\hat{\bf n}) =\sum_{\ell=2}^{\ell_{\rm max}} \tilde{S}_\ell (\hat{\bf n}).
\end{equation}

It is convenient to redefine the parity estimator as $S(\hat{\bf n}) = \tilde{S}_{\rm tot}(\hat{\bf n}) - (\ell_{\rm max}-1)$, so that $\left<S\right> = 0$.

The most even and odd mirror-parity directions for a given map can be considered by estimating \citep{david2012}:
\begin{equation}S_+ = \frac{\max(S)-\mu(S)}{\sigma(S)},\end{equation}
\begin{equation}S_- = \frac{|\min(S)-\mu(S)|}{\sigma(S)},\end{equation}
where $\mu(S)$ and $\sigma(S)$ are the mean and standard deviation of the $S$ map.
The quantities $S_+$ and $S_-$ each correspond to different axes $\hat{\bf n}_+$ and $\hat{\bf n}_-$ respectively and are the quantities which we consider when discussing mirror parity in CMB maps.

{ In order to use inpainted maps to study mirror parity, it is crucial that the inpainting method  constitutes a bias-free reconstruction method, i.e. that it does not destroy existing mirror parities nor that it creates previously inexistent mirror parity anomalies. }

{ In order to test this, we consider three sets of 1000 simulated CMB maps using a WMAP7 best fit cosmology with $nside=32$, one set for each inpainting prior. For each simulation we calculate the mirror parity estimators $S_\pm$. We isolate the anomalous maps, defined as those whose $S_\pm$ value is larger than twice the standard deviation given by the simulations. This gives us $34$ maps ($3.4\%$) for the even mirror parity and $40$ maps ($4\%$) for the odd mirror parity in the full-sky simulations. } 

\begin{figure*}[htb]
\vbox{
\hbox{
\includegraphics [trim= 2cm 6cm 2.cm  4cm , clip,  width=0.75\textwidth]{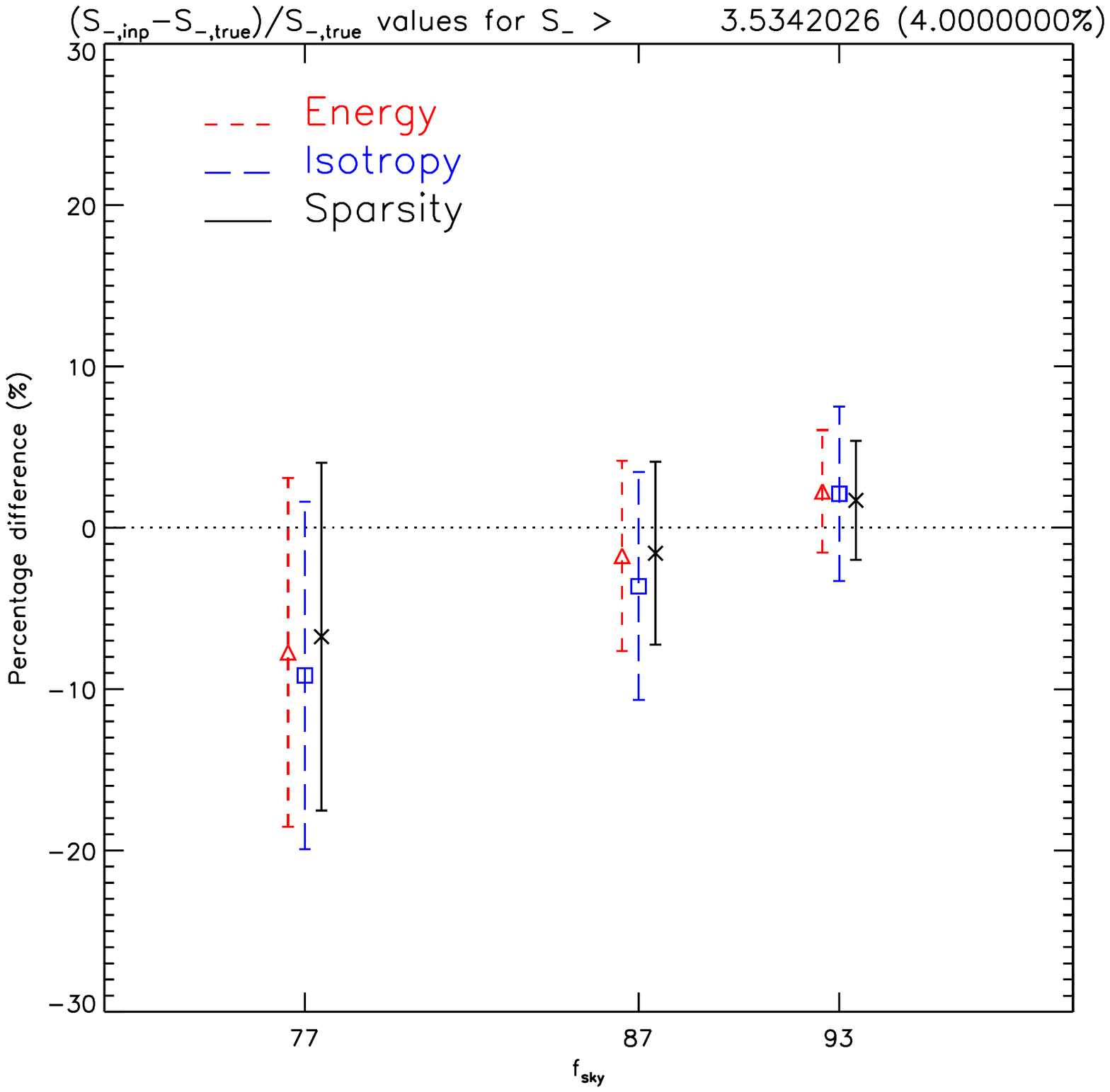}}
}
\caption{Percentage difference between anomalous (i.e. high) odd mirror parity scores ($S_-$) before (i.e. full-sky) and after inpainting for three different inpainting priors.}
\label{fig_anomalies}
\end{figure*}

{ Since recent work shows a potential odd-mirror anomaly in WMAP data \citep{david2012}, we focus on testing for potential biases in odd-mirror anomalies. Fig.\ref{fig_anomalies} show the percentage difference between the true $S_-$ value of the one estimated from the inpainted map. For this statistical test, the result is not dependent on the inpainting method as the three different priors give similar biases and error bars. For positive-mirror anomalies, we find similar results (not shown in Figure).}  
Similarly to the previous experiment, 
we can see that a very good estimation of the parity statistic can be achieved for masks with Fsky larger $0.8$.

\section{Software and Reproducible Research}
To support reproducible research, the developed IDL code will be released with the next version 
of ISAP (Interactive Sparse astronomical data Analysis Packages) via the web site:\\ 
{\centerline{\texttt{http://www.cosmostat.org}}}\\

All the experiments were performed using the default options:
\begin{itemize}
\item{Sparsity:} 
\begin{verbatim}
Alm = cmb_lowl_alm_inpainting(CMBMap,
 Mask, /sparsity)
\end{verbatim}
\item{Energy:} 
\begin{verbatim}
Pd = mrs_powspec( CMBMap,  Mask)
P = mrs_deconv_powspec( pd,  Mask)
Alm = cmb_lowl_alm_inpainting(CMBMap, 
 /Energy, Prea=P)
\end{verbatim}
\item{Isotropy:} 
\begin{verbatim}
 Pd = mrs_powspec( CMBMap,  Mask)
 P = mrs_deconv_powspec( Pd,  Mask)
 Alm = cmb_lowl_alm_inpainting(CMBMap,
  /Isotropy, Prea=P)
\end{verbatim}
\end{itemize}

\section{Conclusion}
We have investigated three priors to regularize the CMB inpainting problem: Gaussianity, sparsity and isotropy. To solve the corresponding minimization problems, we have also proposed fast novel algorithms, based for instance on proximal splitting methods for convex non-smooth optimization. We found that both Gaussianity and $\ell_1$ sparsity priors lead to very good results. The isotropy prior does not provide as good results. The sparsity prior seems to lead to slightly better for $\ell > 2$, and more robust when the sky coverage decreases. Furthermore, unlike the energy-prior based inpainting, the sparsity-based one does not require a power spectrum as an input, which is a significant advantage.

Then we evaluated the reconstruction quality as a function of the sky coverage, and we have seen that a high quality reconstruction, within 1\% of the cosmic variance, can be reached for a mask with $\fsky$ larger than 80\%. {We also studied mirror-parity anomalies, and found that mask with $\fsky$ larger than 80\% also permitted nearly bias-free reconstructions of the mirror parity scores.} Such a mask seems unrealistic for WMAP data analysis, but it could and should be a target for Planck component separation methods. To know if this  80\% mask is realistic for Planck, we will need
more realistic simulations including ISW effect, and residual foregrounds at amplitudes similar to those achieved by the 
actual best Planck component separation methods. Another interesting study will consist in comparing the sparse inpainting to other 
methods such as the maximum likelihood or the Wiener filtering \citep{peiris2011,david2012}.

  
\section*{Acknowledgment}
The authors would like to thank Hiranya Peiris and Stephen Feeney for useful discussions.
This work was supported by the French National Agency for Research (ANR -08-EMER-009-01),  
the European Research Council grant SparseAstro (ERC-228261), and the Swiss National Science Foundation (SNSF).


\section*{Appendix A: Algorithm for the $l_0$ problem}
Solving \eqref{eq:minsparse} when $q=0$ is known to be NP-hard. This is further complicated by the presence of the non-smooth constraint term. Iterative Hard thresholding (IHT) attempts to solve this problem through the following scheme
\begin{equation}
   \cmba^{n+1} =  \Delta^{\mathrm{H}}_{\lambda_n}(\cmba^{n} + {\sht} (\obs - \mask {\sht}^* \cmba^n)) ~,
\label{eqn_mca}
\end{equation}
where the nonlinear operator $\Delta^{\mathrm{H}}_{\lambda}$ is a term-by-term hard thresholding, i.e. $\Delta^{\mathrm{H}}_\lambda(\cmba) = (\rho_\lambda^{\mathrm{H}}(\cmba_i))_i$, where $\rho_\lambda^{\mathrm{H}}(\cmba_i) = \cmba_i$ if $ | \cmba_i | > \lambda$ and $0$ otherwise.
The threshold parameter $\lambda_n$ decreases with the iteration number and is supposed to play a role similar to the cooling parameter in simulated annealing, i.e. hopefully it allows to avoid stationary points and local minima.

\section*{Appendix B: Algorithm for the $l_1$ problem}
It is now well-known that $l_1$ norm is the tightest convex relaxation (in the $\ell_2$ ball) of the $l_0$ penalty \citep{starck:book10}. This suggests solving \eqref{eq:minsparse} with $q=1$. In this case, the problem is well-posed: it has at least a minimizer and all minimizers are global. Furthermore, although it is completely non-smooth, it can be solved efficiently with a provably convergent algorithm belonging to the family of proximal splitting schemes \citep{CombettesPesquet09,starck:book10}. 

In particular, we propose to use the Douglas-Rachford (DR) splitting scheme. Let $\beta > 0$, $\parenth{\alpha_n}_{n \in \NN}$ be a sequence in $]0,2[$ such that $\sum_{t\in\NN} \alpha_n(2-\alpha_n) = +\infty$. The DR recursion, applied to \eqref{eq:minsparse} with $q=1$, reads 
\begin{align}
\label{eq:drl1}
\cmba^{n+\frac{1}{2}} &= \proj_{\{\cmba: \obs =  \mask {\sht}^* \cmba\}}\parenth{\cmba^{n}}, \nonumber \\
\cmba^{n+1}   &= \cmba^{n}  
+ \alpha_n \parenth{\prox_{\beta \norm{\cdot}_1} \parenth{2\cmba^{n+\frac{1}{2}}  - \cmba^{n}} - \cmba^{n+\frac{1}{2}}} ~.
\end{align}
$\prox_{\beta \norm{\cdot}_1}$ is the proximity operator of the $l_1$-norm which can be easily shown to be soft-thresholding
\begin{equation}
\label{eq:proxl1}
\prox_{\beta \norm{\cdot}_1}(\cmba) = \Delta^{\mathrm{S}}_{\beta}(\cmba),
\end{equation}
where $\Delta^\mathrm{S}_\beta(\cmba)=(\rho^\mathrm{S}_\beta(\cmba_i))_i$ and $\rho_\lambda^{\mathrm{S}}(\cmba_i) = \mathrm{sign}(\cmba_i) \mathrm{max}(0, | \cmba_i |  - \beta)$. $\proj_{\{\cmba: \obs =  \mask {\sht}^* \cmba\}}$ is the orthogonal projector on the corresponding set, and it consists in taking the inverse spherical harmonic transform of its argument, setting its pixel values to the observed ones at the corresponding locations, and taking the forward spherical harmonic transform.
It can be shown that the sequence $(\cmba^{n+\frac{1}{2}})n \in \NN_{}$ converges to a global minimizer of \eqref{eq:minsparse} for $q=1$.\\

\medskip

\section*{Appendix C: Algorithm for  the $l_2$ problem, }
This scheme is based again on Douglas-Rachford splitting applied to solve problem \eqref{eq:minwl2}. Its steps are
\begin{align}
\label{eq:drwl2}
\cmb^{n+\frac{1}{2}} &= \proj_{\{\cmb: \obs = \mask \cmb\}}\parenth{\cmb^{n}}, \nonumber \\
\cmb^{n+1}   &= \cmb^{n} + \alpha_n \bigg(\prox_{\beta\norm{\sht(\cdot)}^2_{C^{-1}}} \big(2\cmb^{n+\frac{1}{2}} \nonumber \\
&  \qquad - \cmb^{n}\big) - \cmb^{n+\frac{1}{2}}\bigg) ~,
\end{align}
where $\beta$ and $\alpha_n$ are defined as above, and the proximity operator of the squared weighted $\ell_2$-norm is
\begin{equation}
\label{eq:proxwl2}
\prox_{\beta\norm{\sht(\cdot)}^2_{C^{-1}}}(\cmb) = \sht^*\parenth{(\sht \cmb)  \otimes \parenth{\frac{C }{ \beta+C}}},
\end{equation}
where $\otimes$ stands for the entry-wise multiplication between two vectors, and the division is also to be understood entry-wise. The projector $\proj_{\{\cmb: \obs = \mask \cmb\}}$ has a simple closed-form and consists in setting pixel values to the observed ones at the corresponding locations, and keeping the others intact. It can be shown that the sequence $(\cmb^{n+\frac{1}{2}})_{n \in \NN}$ converges to the unique global minimizer of \eqref{eq:minwl2}.

\section*{Appendix D: Algorithm  for Isotropy Inpainting}
The isotropy-constrained inpainting problem can be cast as the \textit{non-convex} feasibility problem
\begin{equation}
\label{eq:isoinapint}
\text{find} ~ \widehat{\cmb} \in \mathcal{C}_{\veps} \cap \{\cmb: \obs = \mask \cmb\} ~.
\end{equation}
The feasible set is nonempty since the constraint set is non-empty and closed; the CMB is in it under the isotropy hypothesis. To solve \eqref{eq:isoinapint}, we propose to use the von Neumann's method of alternating projections onto the two constraint sets whose recursion can be written
\begin{equation}
\cmb^{(n+1)} = \proj_{\{\cmb: \obs = \mask \cmb\}}\parenth{\proj_{\cC_\veps}\parenth{\cmb^{(n)}}} ~.
\end{equation}
Using closedness and prox-regularity of the constraints, and by arguments from \citet{LewisMalick07}, we can conclude that our non-convex alternating projections algorithm for inpainting converges locally to a point of the intersection $\mathcal{C}_{\veps} \cap \{\cmb: \obs = \mask \cmb\}$ which is non-empty.

\medskip

\bibliographystyle{aa}
\bibliography{JLSBibTex}

\end{document}

%% file: prelim.tex
\newcommand{\vmu}{{\boldsymbol{\mu}}}

\newcommand{\RR}{\mathbb{R}}
\newcommand{\CC}{\mathbb{C}}

\newcommand{\NN}{\mathbb{N}}

\newcommand{\be}{\begin{eqnarray}}
\newcommand{\ee}{\end{eqnarray}}

\newcommand{\st}{\mathrm{\quad s.t. \quad}}

\newcommand{\GN}{\mathcal{N}}

\newcommand{\E}[1]{\mathbb{E}\left[{#1}\right]}

\newcommand{\norm}[1]{\left\|#1\right\|}

\newcommand{\abs}[1]{\left|#1\right|}

\newcommand{\veps}{{\varepsilon}}

\newcommand{\parenth}[1]{\left({#1}\right)}

\newcommand{\ens}[1]{\left\{{#1}\right\}}

\newcommand{\cC}{{\mathcal C}}

\newcommand{\beqn}{\begin{eqnarray}}
\newcommand{\eeqn}{\end{eqnarray}}

\renewcommand{\leq}{\leqslant}

\DeclareMathOperator{\prox}{prox}

\DeclareMathOperator{\proj}{Proj}

\newcommand{\argmin}[1]{\underset{#1}{\mathrm{argmin}}\;}

\newcommand{\Argmin}[1]{\underset{#1}{\mathrm{Argmin}}\;}